\documentclass[10pt,english,british,tightenlines,eqsecnum,
floats,aps,amsmath,amssymb,nofootinbib,superscriptaddress,prd,showpacs,showkeys]
              {revtex4}
\usepackage[latin9]{inputenc}
\usepackage{color}
\usepackage{babel}
\usepackage{amsmath}
\usepackage{amssymb}
\usepackage{graphicx}
\usepackage{esint}
\usepackage[unicode=true,pdfusetitle,
 bookmarks=true,bookmarksnumbered=false,bookmarksopen=false,
 breaklinks=false,pdfborder={0 0 1},backref=false,colorlinks=false]
 {hyperref}
 \makeatletter
\@ifundefined{textcolor}{}
{%
 \definecolor{BLACK}{gray}{0}
 \definecolor{WHITE}{gray}{1}
 \definecolor{RED}{rgb}{1,0,0}
 \definecolor{GREEN}{rgb}{0,1,0}
 \definecolor{BLUE}{rgb}{0,0,1}
 \definecolor{CYAN}{cmyk}{1,0,0,0}
 \definecolor{MAGENTA}{cmyk}{0,1,0,0}
 \definecolor{YELLOW}{cmyk}{0,0,1,0}
}

\@ifundefined{date}{}{\date{09-Feb-2019}}
\usepackage{babel}\usepackage{euscript}\usepackage{epsfig}

\usepackage{amsfonts}

\usepackage{fancyhdr}

\makeatother

\begin{document}
\title{Extended anisotropic models in noncompact Kaluza-Klein theory}

\author{S. M. M. Rasouli}

\email{mrasouli@ubi.pt}

\affiliation{Departamento de F\'{i}sica, Universidade da Beira Interior, Rua Marqu\^{e}s d'Avila
e Bolama, 6200 Covilh\~{a}, Portugal}

\affiliation{Centro de Matem\'{a}tica e Aplica\c{c}\~{o}es (CMA - UBI),
Universidade da Beira Interior, Rua Marqu\^{e}s d'Avila
e Bolama, 6200 Covilh\~{a}, Portugal}



\author{P. V. Moniz}

\email{pmoniz@ubi.pt}

\affiliation{Departamento de F\'{i}sica, Universidade da Beira Interior, Rua Marqu\^{e}s d'Avila
e Bolama, 6200 Covilh\~{a}, Portugal}
\affiliation{Centro de Matem\'{a}tica e Aplica\c{c}\~{o}es (CMA - UBI),
Universidade da Beira Interior, Rua Marqu\^{e}s d'Avila
e Bolama, 6200 Covilh\~{a}, Portugal}
\affiliation{DAMTP, Centre for
  Mathematical Sciences, University of Cambridge, Wilberforce Road, Cambridge CB3 0WA, UK}
\begin{abstract}
In this paper, new exact solutions for locally rotational
symmetric (LRS) space-times are obtained within the modified Brans-Dicke
theory (MBDT) \cite{RFM14}. Specifically, extended
five-dimensional (5D) versions of Kantowski-Sachs, LRS Bianchi type I and Bianchi
type III are investigated in the context of the standard Brans-Dicke theory. We subsequently extract their
corresponding dynamics on a 4D hypersurface.
Our results are discussed regarding others obtained in the standard
Brans-Dicke theory, induced-matter theory and general relativity.
 Moreover, we comment on the evolution of the scale
 factor of the extra spatial dimension, which is of interest
 in Kaluza-Klein frameworks.
 \end{abstract}

\medskip


\maketitle

\section{Introduction}
\label{int} \indent

Several significant results have motivated researchers to consider scalar-tensor
theories (instead of general relativity) in noncompact Kaluza-Klein frameworks, e.g., space-time
matter theory \cite{PW92,stm99,RS99-1,RS99-2}, in order to
establish modified theories~\cite{ARB07,qiang2005,qiang2009,Ponce1,RFM14, RM18}.
In induced-matter theory, by starting from pure geometrical $5D$ field
equations in vacuum, induced (effective) matter is obtained on a $4D$ space time, as a direct
consequence of the extra dimensions, in accordance with
embedding theorems \cite{C26,M63,AL01,ADLR01}.
Moreover, there are corresponding reduced
cosmologies which can bear a reasonable agreement
with specific cosmological observables \cite{R02,W08,NRJ09,RJ10}.
Different cosmological models are present in the
literature~\cite{Ponce1,RFM14,BMM15,MO16,T17,RB18,RM18}, whose (effective)
matter sources emerge from the geometry of
higher dimensions. Such contribution
not only can yield implications similar to ordinary matter sources but it can also play the role of dark
energy or dark matter in the universe \cite{Ponce1,RM16,B17,RM18}.

Within MBDT \cite{RFM14} (as well as other noncompact Kaluza-Klein
models \cite{AB04}) a significant advantage is achieved.
For instance, to obtain an accelerating scale factor,
a scalar potential does not need to be added to the action by hand~\cite{SS01}, but instead, an
induced potential is dictated from the intrinsic geometry~\cite{RFM14,AB04,RM18} (recently, it has been
shown that appropriate kinetic inflation can be obtained in the absence of a
 cosmological constant, ordinary matter and scalar potential \cite{RFK11, RM14, RSFMM18, RMM18, RM16-U}).
 An alternative approach was proposed in \cite{BG18-2}.
 Concretely, the dynamical space-time theory \cite{BG18} has
 been applied to a particular case of Kaluza-Klein cosmology,
 associated with a torus space (for a detailed study of
 this model, see \cite{T84}), and a mechanism of inflation has been analyzed therein.

 In this paper, we are interested in applying the MBDT in obtaining anisotropic cosmological solutions on a hypersurface.
Let us note that without any restricting conditions (for instance, a cylinder condition on
the extra coordinate and/or a higher dimensional matter hypothesis~\cite{qiang2005,qiang2009}), it
has been shown that the MBDT \cite{RFM14}
 possesses four sets of field equations established on a $D$-dimensional hypersurface
 orthogonal to the extra dimension: two sets correspond to those of the
standard  Brans-Dicke (BD) theory \cite{BD61}, including a self interacting scalar potential; one set can be considered as an extended
 conservation law introduced in the induced-matter theory \cite{PW92}; the other one has
 no counterpart in either standard BD theory or in the induced-matter theory.
 Furthermore, the effective matter as well as the induced
 scalar potential introduced in~\cite{RFM14} emerge entirely from the
 geometry of the extra dimension.

 A generalized Friedmann-Robertson-Walker (FRW)
 universe for the three values of the spatial curvature index has been investigated in~\cite{RM16}.
  However, our universe at very early times may not have been so completely uniform.
In this respect, a generalized Bianchi type I anisotropic universe has also been
 studied in the context of the induced-matter theory and MBDT~\cite{H00,PW08,P09,RFS11}.
Therefore, the purpose of our herein work is to obtain new exact
 solutions for the field equations of the BD cosmology,
  in which the universe is described by either of three different anisotropic
  space times. More precisely, we will consider the extended
  versions of Kantowski-Sachs~\cite{KS66,BD97}, LRS Bianchi type I and Bianchi type III line elements in vacuum.
  Subsequently, by applying the MBDT, the corresponding reduced cosmology will be analysed on a $4D$ hypersuface.
 Our new solutions will be compared with some produced from standard
 scalar-tensor theories as well as from general relativity.

Our research also conveys another innovative feature to obtain exact cosmological solutions (which play a
pivotal role in cosmology), providing insights
into the quantitative as well as qualitative behaviour of our universe.
Specifically, one of the advantages of the modified induced-matter models
(based on the scalar-tensor theories \cite{RM18,RFM14,PW08})
is that the methodology may assist in solving the field equations more easier. Let us be more precise.
 In the herein framework, we benefit from solving the field equations in the bulk in
 the absence of an energy momentum tensor. Subsequently, we construct the corresponding dynamics on the
 hypersurface.
 Moreover, in contrast to phenomenological frameworks, we do not assume a scalar potential (on the hypersurface) for obtaining the
  favorable consequences. Instead, an induced scalar potential is dictated from a geometrical reduction procedure.

Our paper is organized as follows. In Section~\ref{Set up}, we
present a brief review of the MBDT.
In Section~\ref{bulk}, we consider
a $5D$ vacuum universe, which is described by
Kantowski-Sachs, Bianchi type I and Bianchi type
III line-elements. By defining a new time coordinate, we
 find new exact cosmological solutions for the field equations.
Moreover, we obtain a set of constraints for the
 parameters of the models and we investigate particular cases of these solutions.
 In section~\ref{reduced cosmology}, we derive the
 effective energy momentum tensor, induced scalar potential and other
 physical quantities associated with our herein anisotropic models. Then, we study a corresponding
 reduced cosmology in greater detail.
 We show that the conservation law for the resulted induced matter is satisfied identically for all the models.
In section \ref{Deparametrized}, when it is possible, the solutions are
represented in terms of the cosmic time. For the models whose solutions
are not feasible to represent in terms of the cosmic time, we restrict
ourselves to analyse the consequences in terms of the conformal time.
Finally, we present a summary and conclusions in Section~\ref{conclusion}.

\section{Modified Brans-Dicke theory in four dimensions}
\label{Set up}
\indent
In this section, let us provide a brief review of the MBDT~\cite{RFM14}.
The $5D$ action of the BD theory in the Jordan frame can be given by~\cite{RFM14,RM16}
\begin{equation}\label{(D+1)-action}
{\cal S}^{(5)}=\int d^{5}x \sqrt{\Bigl|{}{\cal
G}^{(5)}\Bigr|} \,\left[\phi
R^{(5)}-\frac{\omega}{\phi}\, {\cal
G}^{ab}\,(\nabla_a\phi)(\nabla_b\phi)+16\pi\,
L\!^{(5)}_{_{\rm matt}}\right],
\end{equation}
where $\omega$ and $\phi$ are the BD coupling parameter and
BD scalar field, respectively; the Latin indices take values from $0$ to $4$ and
 $L^{(5)}_{\rm matt}$ is the Lagrangian density of the ordinary matter (in five dimensions).
The determinant of the $5D$ metric ${\cal G}_{ab}$ is denoted
by ${\cal G}^{(5)}$; $R^{(5)}$ is the Ricci curvature scalar and $\nabla_a$ represents
the covariant derivative in the $5D$ space-time.
Throughout this paper, we use Planck units.

The field equations extracted from the action (\ref{(D+1)-action}) can be written as
\begin{equation}\label{(D+1)-equation-1}
G^{(5)}_{ab}=\frac{8\pi}{\phi}\,T^{(5)}_{ab}+\frac{\omega}{\phi^{2}}
\left[(\nabla_a\phi)(\nabla_b\phi)-\frac{1}{2}{\cal G}_{ab}(\nabla^c\phi)(\nabla_c\phi)\right]
+\frac{1}{\phi}\Big(\nabla_a\nabla_b\phi-{\cal G}_{ab}\nabla^2\phi\Big)
\end{equation}
and
\begin{equation}\label{(D+1)-equation-4}
\nabla^2\phi=\frac{8\pi T^{(5)}}{3\omega+4},
\end{equation}
where $\nabla^2\equiv\nabla_a\nabla^a$ and $T^{(5)}={\cal G}^{ab}T^{(5)}_{ab}$ is the trace
of the energy-momentum tensor $T^{(5)}_{ab}$ associated with the ordinary matter fields in a
$5D$ space-time.

The field equations of MBDT convey the dynamics on the $4D$ hypersurface \cite{RFM14}.
More specifically, we take the line-element \cite{OW97}
\begin{equation}\label{global-metric}
dS^{2}={\cal G}_{ab}(x^c)dx^{a}dx^{b}=
g_{\mu\nu}(x^\alpha,l)dx^{\mu}dx^{\nu}+
\epsilon\psi^2\left(x^\alpha,l\right)dl^{2}.
\end{equation}
We use the notation $x^\alpha=(x^0,x^1,x^2,x^3)$ for
the coordinates in $4D$ space-time and $l$ is the
non-compact coordinate associated with the fifth dimension.
Moreover, we have $\epsilon=\pm1$ to indicate if the extra dimension is either time-like or space-like.
We are also assuming a specific hypersurface $\Sigma_0 (l=l_0={\rm constant}$), which is orthogonal to
the unit vector $n^a=\delta^a_{_4}/\psi$ (where $n_an^a=\epsilon$).
Among four sets of modified field equations associated with the MBDT, only
the following two sets are of interest for our objectives in this paper:
\begin{eqnarray}\nonumber
G_{\mu\nu}^{(4)}&=&\frac{8\pi}{\phi}\,
\left(S_{\mu\nu}+T_{\mu\nu}^{^{[\rm BD]}}\right)
+
\frac{\omega}{\phi^2}\left[({\cal D}_\mu\phi)({\cal D}_\nu\phi)-
\frac{1}{2}g_{\mu\nu}({\cal D}_\alpha\phi)({\cal
D}^\alpha\phi)\right]\\ \nonumber\\
 &&+
 \frac{1}{\phi}\left({\cal D}_\mu{\cal
D}_\nu\phi- g_{\mu\nu}{\cal
D}^2\phi\right)
-g_{\mu\nu}\frac{V(\phi)}{2\phi} \label{BD-Eq-DD}
\end{eqnarray}
and
\begin{eqnarray}\label{D2-phi}
{\cal D}^2\phi=\frac{8\pi}{2\omega+3}\left(S+T^{^{[\rm BD]}}\right)+
\frac{1}{2\omega+3}\left[\phi\frac{dV(\phi)}{d\phi}-2V(\phi)\right].
\end{eqnarray}
In what follows let us briefly explain
the symbols and quantities that appear in equations (2.5) and (2.6); for detailed review of the
MBDT, see \cite{RFM14}.
In these equations, ${\cal D}_\alpha$ denotes the covariant derivative on a $4D$
hypersurface and ${\cal D}^2\equiv{\cal D}_\alpha{\cal D}^\alpha$. In (\ref{BD-Eq-DD}) we have defined
\begin{eqnarray}\label{S}
S_{\mu\nu}\equiv T_{\mu\nu}^{(5)}-
g_{\mu\nu}\left[\frac{(\omega+1)T^{(5)}}{3\omega+4}-
\frac{\epsilon\, T_{44}^{(5)}}{\psi^2}\right],
\end{eqnarray}
constituting the effective part of ordinary matter that can be assumed in $5D$ bulk. In addition, we also introduced
\begin{eqnarray}\label{matt.def}
T_{\mu\nu}^{^{[\rm BD]}}\equiv T_{\mu\nu}^{^{[\rm IMT]}}+T_{\mu\nu}^{^{[\rm \phi]}}
+\frac{1}{16\pi}g_{\mu\nu}V(\phi),
\end{eqnarray}
which is an induced geometrically energy momentum tensor, as is composed of the following parts
\begin{eqnarray}\label{IMTmatt.def}
\frac{8\pi}{\phi}T_{\mu\nu}^{^{[\rm IMT]}}\!&\equiv&\!
\frac{{\cal D}_\mu{\cal D}_\nu\psi}{\psi}
-\frac{\epsilon}{2\psi^{2}}\left(\frac{\overset{*}{\psi}
\overset{*}{g}_{\mu\nu}}{\psi}-\overset{**}{g}_{\mu\nu}
+g^{\lambda\alpha}\overset{*}{g}_{\mu\lambda}\overset{*}{g}_{\nu\alpha}
-\frac{1}{2}g^{\alpha\beta}\overset{*}{g}_{\alpha\beta}\overset{*}{g}_{\mu\nu}\right)\cr
 \!\!\!&&\!\!-\frac{\epsilon g_{\mu\nu}}{8\psi^2}
\left[\overset{*}{g}^{\alpha\beta}\overset{*}{g}_{\alpha\beta}
+\left(g^{\alpha\beta}\overset{*}{g}_{\alpha\beta}\right)^{2}\right],\\\nonumber
\\
\label{T-phi} \frac{8\pi}{\phi}T_{\mu\nu}^{^{[\rm
\phi]}}\!&\equiv &\!
\frac{\epsilon\overset{*}{\phi}}{2\psi^2\phi}\left[\overset{*}{g}_{\mu\nu}
+g_{\mu\nu}\left(\frac{\omega\overset{*}{\phi}}{\phi}-g^{\alpha\beta}\overset{*}{g}_{\alpha\beta}\right)\right].
\end{eqnarray}
Moreover, the notation $\overset{*}{A}$ has been used to denote the derivative of a quantity $A$  with respect to the fifth coordinate, $l$.
Finally, the induced scalar potential $V(\phi)$ is derived from the following differential equation (see Ref. \cite{RFM14})
\begin{eqnarray}\label{v-def}
\phi \frac{dV(\phi)}{d\phi}\!&\equiv&\!-2(\omega+1)
\left[\frac{({\cal D}_\alpha\psi)({\cal D}^\alpha\phi)}{\psi}
+\frac{\epsilon}{\psi^2}\left(\overset{*}{\phi}-
\frac{\overset{*}{\psi}\overset{*}{\phi}}{\psi}\right)\right]-\frac{2\epsilon\omega\overset{*}{\phi}}{2\psi^2}
\left(\frac{\overset{*}{\phi}}{\phi}+g^{\mu\nu}\overset{*}{g}_{\mu\nu}\right)\\\nonumber
\\\nonumber
&&\!
+\frac{\epsilon\phi}{4\psi^2}
\left[\overset{*}{g}^{\alpha\beta}\overset{*}{g}_{\alpha\beta}
+\left(g^{\alpha\beta}\overset{*}{g}_{\alpha\beta}\right)^2\right]
+16\pi\left[\frac{(\omega+1)T^{(5)}}{3\omega+4}-\frac{\epsilon T_{44}^{(5)}}{\psi^2}\right].
\end{eqnarray}
Let us close this section by mentioning a few points:
(i) in general, the BD scalar field is not a constant and therefore $T_{\mu\nu}^{^{[\rm IMT]}}$ is
a generalized version of the corresponding quantity introduced in~\cite{PW92,stm99};
(ii) $T_{\mu\nu}^{^{[\rm \phi]}}$ is constructed
from $\phi$ and its derivative with respect to $l$ and it has no analogue in the induced-matter theory;
(iii) the scalar potential emerges solely from the geometry of the fifth dimension, rather
than considering any ad hoc phenomenological assumptions \cite{SS01}.
(iv) equations (\ref{BD-Eq-DD}) and (\ref{D2-phi}) can be derived from the following action:
 ${\cal S}^{^{(4)}}\!\!\!\!
=\!\int d^{^{\,4}}\!x \sqrt{-g}\,\left[\phi
R^{^{(4)}}-\frac{\omega}{\phi}\, g^{\alpha\beta}\,({\cal
D}_\alpha\phi)({\cal D}_\beta\phi)-V(\phi)+16\pi\,
L\!^{^{(4)}}_{_{\rm matt}}\right]$,
 where
$\sqrt{-g}\left(S_{\alpha\beta}+T^{^{[\rm
BD]}}_{\alpha\beta}\right)\equiv 2\delta\left( \sqrt{-g}\,
L\!^{^{(4)}}_{_{\rm matt}}\right)/\delta g^{\alpha\beta}$ and $T^{^{[\rm
BD]}}_{\alpha\beta}$ and $S_{\alpha\beta}$ are covariantly conserved.

In section~\ref{reduced cosmology}, we will employ the above geometrical description to
establish the reduced cosmological models associated with the Kantowski-Sachs, LRS
Bianchi type I and Bianchi type III metrics. Moreover,
we will compare our herein results with others obtained instead in the context of the standard
BD theory, induced-matter theory and general relativity.

\section{Exact Brans-Dicke anisotropic vacuum solutions in a five-dimensional space-time}
\label{bulk}
\indent
In this section, we will use the $5D$ field equations to analyse a
vacuum (i.e., $T_{ab}^{(5)}=0$) universe described by the extended versions of the spatially homogeneous
and anisotropic Kantowski-Sachs, LRS Bianchi type I and Bianchi type III space times.
We solve the equations analytically and obtain exact solutions in the bulk.
We assume the line element as~\cite{KS66,BD97}
\begin{eqnarray}\label{DO-metric}
dS^{2}=-dt^{2}+a^{2}(t)dr^2+
b^2(t)d\Omega^2_\zeta+\epsilon \psi^2(t)dl^{2},
\end{eqnarray}
where the angular metric is given by
\begin{eqnarray}\nonumber
d\Omega^2_\zeta&=&d\theta^2+ f^2(\theta)d\phi^2,\\\nonumber\\\label{f}
f(\theta)&=&\left \{\begin{array}{c}
\!\!\!\!\!\!\!\!\!\!\!\!sin\theta,
\hspace{8mm}\zeta=+1 \hspace{8mm} {\rm (Kantowski-Sachs),}\\\\
\!\!\theta, \hspace{12mm}\zeta=0\hspace{10mm} {\rm (LRS\hspace{2mm} Bianchi \hspace{2mm}type \hspace{2mm}I),}\\\\
sinh\theta,
\hspace{7mm}\zeta=-1\hspace{6mm} {\rm (LRS\hspace{2mm} Bianchi \hspace{2mm}type \hspace{2mm}III),}
 \end{array}\right.
\end{eqnarray}
whit $t$ being the cosmic time;
$a(t)$, $b(t)$ and $\psi(t)$ are cosmological scale factors.

From~(\ref{(D+1)-equation-1}), (\ref{(D+1)-equation-4}) and the
line-element (\ref{DO-metric}), we obtain the equations of
 motion associated with all the three curvatures as
\begin{eqnarray}\label{dot-1}
\frac{\ddot{\phi}}{\phi}\!\!&+&\!\!\frac{\dot{\phi}}{\phi}\left(\frac{\dot{a}}{a}+\frac{2\dot{b}}{b}+\frac{\dot{\psi}}{\psi}\right)=0,\\\nonumber\\
\label{dot-2}
\frac{\dot{b}}{b}\left(\frac{2\dot{a}}{a}+\frac{\dot{b}}{b}\right)\!\!&+&\!\!\frac{\dot{\phi}}{\phi}
\left[\frac{\dot{a}}{a}+\frac{2\dot{b}}{b}-\frac{\omega}{2}\left(\frac{\dot{\phi}}{\phi}\right)+\frac{\dot{\psi}}{\psi}\right]
+\frac{\dot{\psi}}{\psi}\left(\frac{\dot{a}}{a}+\frac{2\dot{b}}{b}\right)+\frac{\zeta}{b^2}=0,\\\nonumber\\
\label{dot-3}
\frac{\ddot{b}}{b}\!\!&+&\!\!\frac{\dot{b}}{b}\left(\frac{\dot{a}}{a}+\frac{\dot{b}}{b}+\frac{\dot{\phi}}{\phi}\right)
+\frac{1}{2}\left[\frac{\ddot{\psi}}{\psi}+\frac{\dot{\psi}}{\psi}\left(\frac{\dot{a}}{a}+
\frac{4\dot{b}}{b}+\frac{\dot{\phi}}{\phi}\right)\right]+\frac{\zeta}{b^2}=0,\\\nonumber\\
\label{dot-4}
\frac{\ddot{a}}{a}\!\!&+&\!\!\frac{\dot{a}}{a}\left(\frac{2\dot{b}}{b}+\frac{\dot{\phi}}{\phi}\right)
+\frac{1}{2}\left[\frac{\ddot{\psi}}{\psi}+\frac{\dot{\psi}}{\psi}\left(\frac{3\dot{a}}{a}+
\frac{2\dot{b}}{b}+\frac{\dot{\phi}}{\phi}\right)\right]=0,\\\nonumber\\
\label{dot-5}
\frac{\ddot{a}}{a}\!\!&+&\!\!\frac{2\ddot{b}}{b}+\frac{\dot{b}}{b}\left(\frac{2\dot{a}}{a}+\frac{\dot{b}}{b}\right)
+\frac{\dot{\phi}}{\phi}\left[\frac{\omega}{2}\left(\frac{\dot{\phi}}{\phi}\right)-\frac{\dot{\psi}}{\psi}\right]+\frac{\zeta}{b^2}=0,
\end{eqnarray}
where an overdot represents a derivative
with respect to the cosmic time. Since the metrics are spatially homogeneous, we have taken the BD
scalar field depending only on the cosmic time.

Concerning field
equations~(\ref{dot-1})-(\ref{dot-5}), we should note that there are four
unknowns $a$, $b$, $\phi$ and $\psi$, with five coupled non-linear field
equations which are not independent.
To obtain exact solutions we
introduce a new time coordinate,$\eta$, which is related to the
cosmic time $t$ as (notwithstanding the specific power-law assumption taken
 for the Bianchi type I model in \cite{RFS11}, we will show that using the
following transformation yields more generalized set of solutions;
such a coordinate transformation has also been used in \cite{LP-rev})
\begin{eqnarray}
\label{conformal}
dt=bd\eta.
\end{eqnarray}

Consequently, equations~(\ref{dot-1})-(\ref{dot-5}) in terms of the new time coordinate $\eta$ can be rewritten as
\begin{eqnarray}
\label{prime-1}
ab\psi\phi'&=&c_1,\\\nonumber
\\
\label{prime-2}
Z''+\zeta Z\!&=&\!0,\hspace{10mm} {\rm where}\hspace{10mm}Z\equiv ab\phi\psi,\\\nonumber
\\
\label{prime-3}
(XZ)'+2\zeta Z\!&=&\!0,\hspace{10mm} {\rm where}\hspace{10mm} X\equiv [{\rm ln}(ab^2)]',\\\nonumber
\\
\label{prime-4}
[{\rm ln}(YZ)]'\!&=&\!0,\hspace{10mm} {\rm where}\hspace{10mm}Y\equiv[{\rm ln}(a\psi^{\frac{1}{2}})]',\\\nonumber\\
\label{prime-5}
\frac{b'}{b}\left(\frac{2a'}{a}+\frac{b'}{b}\right)\!&+&\!\frac{\phi'}{\phi}
\left[\frac{\psi'}{\psi}-\frac{\omega}{2}\left(\frac{\phi'}{\phi}\right)\right]
+\left(\frac{a'}{a}+\frac{2b'}{b}\right)\left(\frac{\phi'}{\phi}+\frac{\psi'}{\psi}\right)+\zeta=0,
\end{eqnarray}
with $c_1$ being an integration constant and a prime represents $d/d\eta$.

From~(\ref{prime-2}), we get $Z(\eta)=Z_0f(\eta+\eta_0)$ where $Z_0\neq0$ is an integration constant.
Without loss of generality,
we can set $\eta_0=0$.
It is straightforward to retrieve the exact solutions for the
field equations:
\begin{itemize}
  \item { $\zeta=-1$:}\\
  \begin{eqnarray}
\label{BT-sol-1}
a(\eta)&=&a_0\left[tanh\left(\frac{\eta}{2}\right)\right]^{m_1},\hspace{18mm}
b(\eta)=b_0sinh\eta\left[tanh\left(\frac{\eta}{2}\right)\right]^{m_2},\\\nonumber
\\
\phi(\eta)&=&\phi_0\left[tanh\left(\frac{\eta}{2}\right)\right]^{m_3},\hspace{19mm}
\psi(\eta)=\psi_0\left[tanh\left(\frac{\eta}{2}\right)\right]^{m_4},\label{BT-sol-3}
\end{eqnarray}
\item {\it $\zeta=+1$ :}
  \begin{eqnarray}
\label{sol-1}
a(\eta)&=&a_0\left[tan\left(\frac{\eta}{2}\right)\right]^{m_1},\hspace{18mm}
b(\eta)=b_0sin\eta\left[tan\left(\frac{\eta}{2}\right)\right]^{m_2},
\\\nonumber
\\
\label{sol-3}
\phi(\eta)&=&\phi_0\left[tan\left(\frac{\eta}{2}\right)\right]^{m_3},\hspace{17mm}
\psi(\eta)=\psi_0\left[tan\left(\frac{\eta}{2}\right)\right]^{m_4},
\end{eqnarray}
\item {\it $\zeta=0$ :}
  \begin{eqnarray}
a(\eta)=a_0\eta^{n_1},\hspace{10mm}
b(\eta)=b_0\eta^{n_2},\hspace{10mm}\phi(\eta)&=&\phi_0\eta^{n_3},\hspace{10mm}\psi(\eta)=\psi_0\eta^{n_4},\label{sol-3-BT1}
\end{eqnarray}
where $m_i$ and $n_i$ ($i=1,2,3,4$) are given by
 \begin{eqnarray}\label{c1}
m_1&\equiv&\frac{2}{3}\left(2\alpha+\beta\right), \hspace{30mm}   m_2\equiv-\frac{1}{3}\left(2\alpha+\beta\right),
\\\nonumber\\
\label{c2}
 m_3&\equiv&\beta,\hspace{45mm} m_4\equiv-\frac{2}{3}\left(\alpha+2\beta\right),\\\nonumber\\
\label{c3}
n_1&\equiv&\frac{2}{3}\left[2\alpha+\beta-\frac{1}{2}(\gamma+3)\right],\hspace{12mm}
 n_2\equiv-\frac{1}{3}\left(2\alpha+\beta+\gamma\right),\\\nonumber\\
 \label{c4}
 n_3&\equiv&\beta, \hspace{45mm} n_4\equiv-\frac{2}{3}\left[\alpha+2\beta-(\gamma+3)\right],
 \end{eqnarray}
\end{itemize}
with $a_0$, $b_0$, $\phi_0$, $\psi_0$, $\alpha$, $\gamma$ and $\beta\equiv\frac{c}{Z_0}$ constituting integration
constants or parameters.
We should note that the integration constants are related
 as $a_0b_0\phi_0\psi_0=Z_0$, which is valid for
the models (\ref{f}). However, relating the parameters $\alpha$, $\beta$ (and $\gamma$) for $\zeta=0$ does differ
from the other two models.
More concretely, for $\zeta=\pm1$, from equation~(\ref{prime-5}), we get
 \begin{eqnarray}\label{sol-5}
4\alpha^2&+&6\alpha\beta+\left(\frac{3\omega}{2}+5\right)\beta^2-3=0,
\end{eqnarray}
or
  \begin{eqnarray}\label{omega}
\omega= -\frac{2(4\alpha^2+6\alpha\beta+5\beta^2-3)}{3\beta^2}.
  \end{eqnarray}
  Whereas, for $\zeta=0$, it is instead
\begin{eqnarray}\label{con.BT1}
-8\alpha^2+4\alpha(3-3\beta+\gamma)+6\beta(3+\gamma)-2[6+\gamma(6+\gamma)]-\beta^2(3\omega+10)=0,
  \end{eqnarray}
  which can be rewritten as
  \begin{eqnarray}\label{omega.BT1}
  \omega=\frac{-2}{3\beta^2}\left[4\alpha^2+2\alpha(3\beta-\gamma-3)+5\beta^2-3\beta(\gamma+3)+\gamma^2+6\gamma+6\right],
  \end{eqnarray}
where we made care of equation~(\ref{prime-5}).

Furthermore, in analogy with the Kasner relations in general
relativity, it is straightforward to show that there are constraints
which relate the powers present in (\ref{BT-sol-1})-(\ref{sol-3-BT1})
\begin{eqnarray}\label{Constraints-m-n}
 \sum_{i=1}^4m_i&=&0,\hspace{10mm}\sum_{i=1}^4m_i^2=2-\omega m_3^2,\\
 \sum_{i=1}^4n_i&=&1,\hspace{10mm}\sum_{i=1}^4n_i^2=1+2n_2-\omega n_3^2,
  \end{eqnarray}
where we have imported
(\ref{omega}) and (\ref{omega.BT1}).

\begin{figure}
\centering\includegraphics[width=3in]{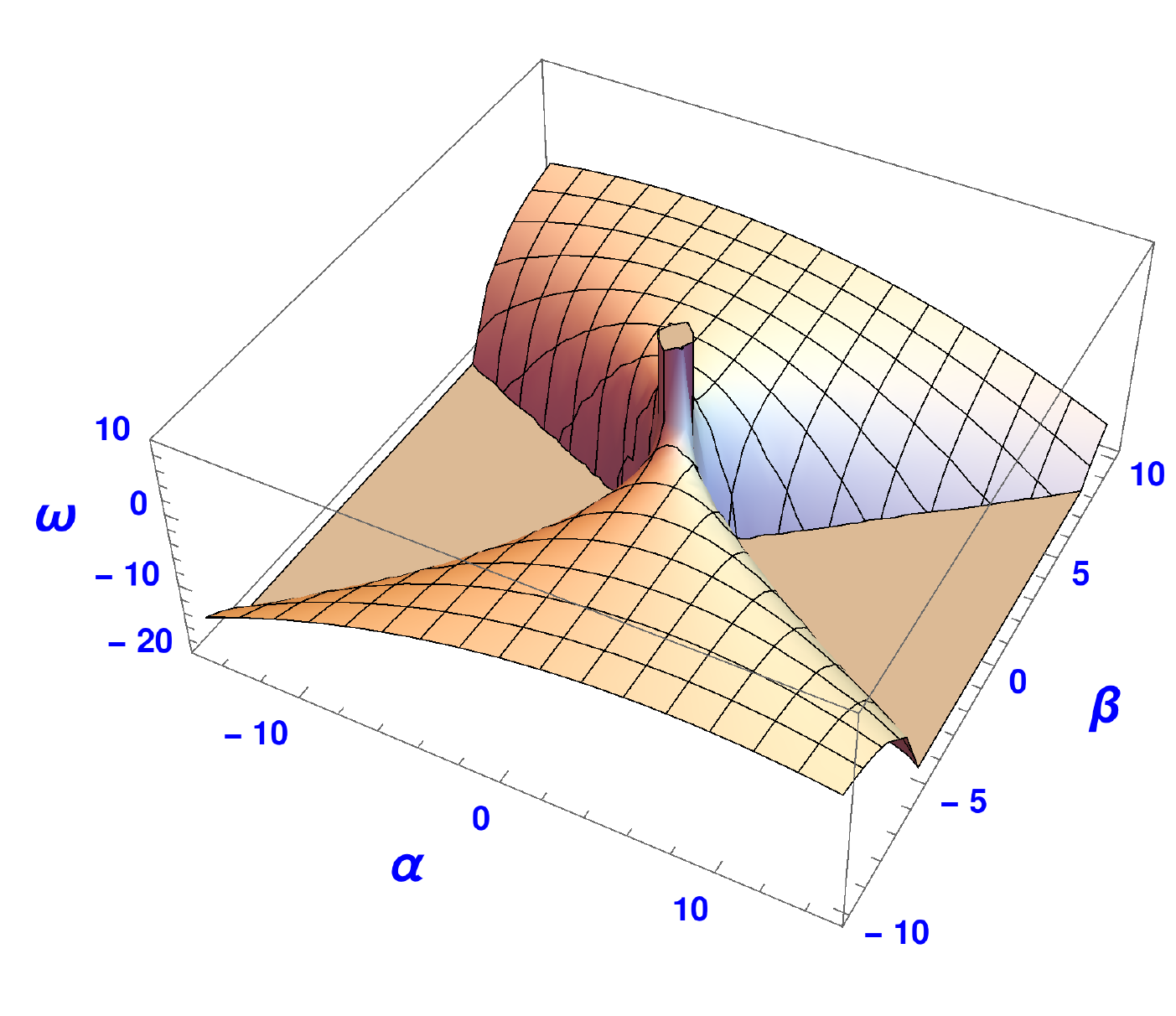}
\hspace{5mm}
\centering\includegraphics[width=3in]{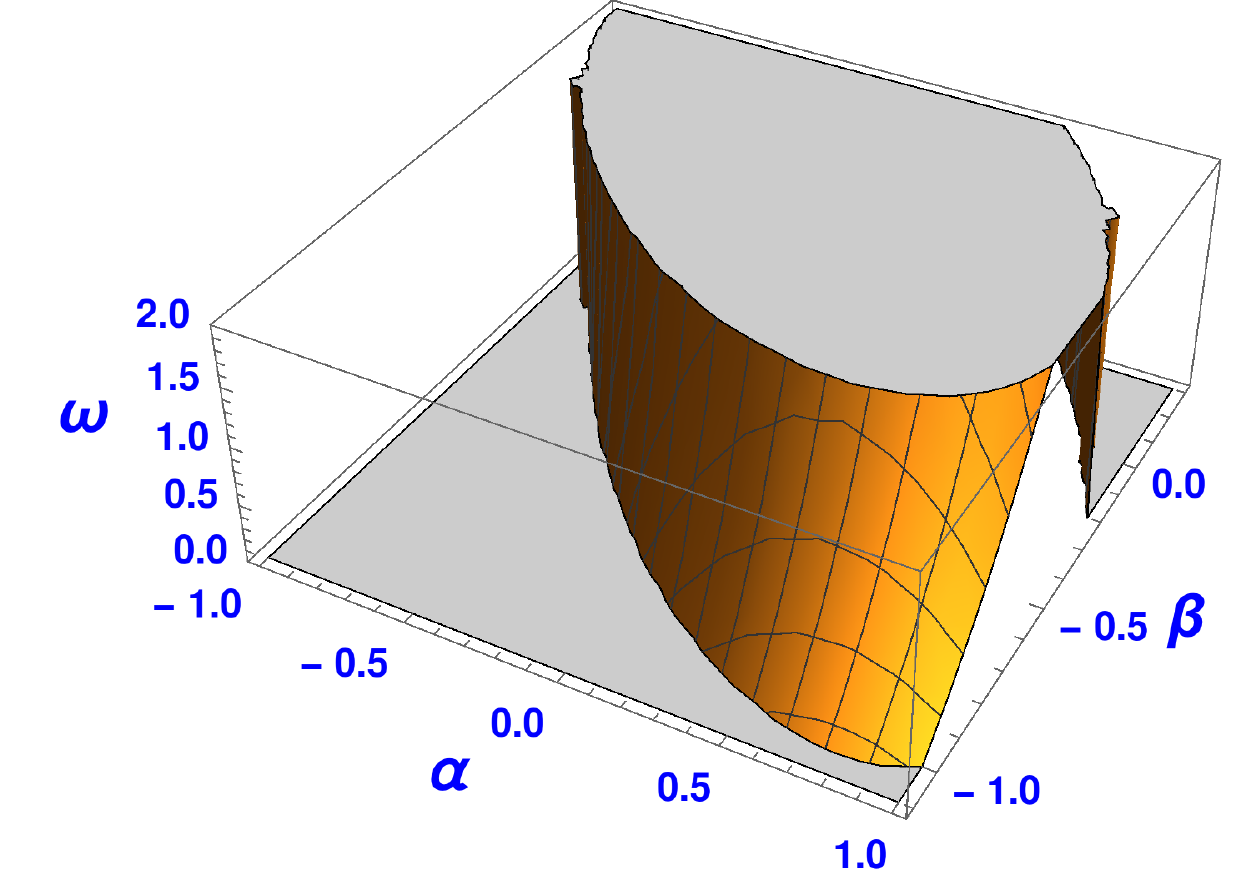}
\caption{The allowed range of $\omega$ (associated with $\zeta=\pm1$) in terms of $\alpha$ and $\beta$.
The right panel shows that for very restricted ranges of $\alpha$ and $\beta$, the BD coupling parameter can take positive values.}
\label{BT3-KS-omega}
\end{figure}

  We see that for the solutions associated with $\zeta=\pm1$, we have just two independent
  parameters and the third one is constrained by relation (\ref{sol-5}).
  It is worthwhile to plot $\omega$ in terms of $\alpha$
  and $\beta$ for
 $\zeta=\pm1$, see figure~\ref{BT3-KS-omega}.
  As seen from the right panel, for restricted intervals
  of $\alpha$ and $\beta$, it is possible to get positive values
  for $\omega$. Concerning the allowed range of the
  BD coupling parameter for the Bianchi type I, we will discuss it with the
  deparametrized solutions in section \ref{Deparametrized}.


We should mention a few particular cases:
 (i) For all three models, when $\beta$ goes to zero, then $|\omega|$ tends to infinity and consequently we
get $\phi=\phi_0={\rm constant}$; therefore the
solutions (\ref{BT-sol-1})-(\ref{sol-3}) may reduce to
those derived in a $5D$ vacuum space time in the context of general relativity (we should mention that when $\omega$ tends to
infinity, the BD solutions reduce (but not always~\cite{BR93,BS97,Faraoni99}) to the corresponding ones in general relativity).
(ii) By assuming $\alpha=-2\beta$ and $\alpha+2\beta=\gamma+3$, associated
with $\zeta=\pm1$ and $\zeta=0$, respectively, then $\psi(\eta)$ takes constant values and
our solutions may reduce to those obtained in the context of the BD cosmology in
 a $4D$ vacuum space time.
(iii) For $\zeta=0$ and $\zeta=+1$, when $a=b$, the solutions may reduce to the corresponding ones
obtained for the spatially flat and closed FRW universes, respectively, in the context of the BD theory.

\section{Effective Brans-Dicke cosmologies on a four dimensional hypersurface}
\label{reduced cosmology}
In the present section, by means of the framework reviewed in section~\ref{Set up},
we will obtain the components of the effective energy momentum tensor and
the induced scalar potential. These will then assist us to retrieve the dynamics on the $4D$ hypersuface.

Substituting the components of the metric~(\ref{DO-metric}) to (\ref{matt.def}), it is straightforward
 to show that the non-vanishing components, in terms of the comoving time,
on the hypersuface, are given by
\begin{eqnarray}\label{t-00}
\frac{8\pi}{\phi}T^{0[{\rm BD}]}_{\,\,\,0}\!&=&\!
-\frac{\ddot{\psi}}{\psi}+\frac{V(\phi)}{2\phi},\\\nonumber
\\
\label{t-ii}
\frac{8\pi}{\phi}T^{1[{\rm BD}]}_{\,\,\,1}\!&=&\!
-\frac{\dot{a}\dot{\psi}}{a\psi}+\frac{V(\phi)}{2\phi},
\end{eqnarray}
where the induced scalar potential $V(\phi)$ is obtained from~(\ref{v-def}).
Note that when we replace $a$ by $b$ in relation~(\ref{t-ii}), then
we get $\frac{8\pi}{\phi}T^{2[{\rm BD}]}_{\,\,\,2}$ (which is equal to $\frac{8\pi}{\phi}T^{3[{\rm BD}]}_{\,\,\,3}$) for
any of the models in (\ref{f}).

In terms of new time coordinate $\eta$, the components of the induced matter are:
\begin{eqnarray}\label{t-00-eta-general}
\rho(\eta)\equiv-T^{0[{\rm BD}]}_{\,\,\,0}\!&=&\!
\frac{\phi(\eta)}{8\pi b^2(\eta)}\left(\frac{\psi''}{\psi}-\frac{b'\psi'}{b\psi}\right)-\frac{V(\eta)}{16\pi},\\\nonumber
\\\nonumber
\\
\label{t-11-eta-general}
P_1(\eta)\equiv T^{1[{\rm BD}]}_{\,\,\,1}\!&=&\!
-\frac{\phi(\eta)}{8\pi b^2(\eta)}\frac{a'}{a}\frac{\psi'}{\psi}+\frac{V(\eta)}{16\pi},\\\nonumber
\\\nonumber
\\
\label{t-22-eta-general}
P_2(\eta)\equiv T^{2[{\rm BD}]}_{\,\,\,2}\!&=&\!P_3(\eta)\equiv T^{3[{\rm BD}]}_{\,\,\,3}=
-\frac{\phi(\eta)}{8\pi b^2(\eta)}\frac{b'}{b}\frac{\psi'}{\psi}+\frac{V(\eta)}{16\pi}.
\end{eqnarray}
Moreover, equation~(\ref{v-def}) becomes
\begin{equation}\label{pot-eta-all}
\frac{dV(\phi)}{d\phi}=\frac{2(1+\omega)}{b^2(\eta)}\left(\frac{\phi'}{\phi}\right)\left(\frac{\psi'}{\psi}\right).
\end{equation}


\subsection{Effective cosmologies for $\zeta=\pm1$}
In order to obtain the components of the induced energy momentum tensor, let us
first calculate the expression for the induced scalar potential.
Substituting solutions (\ref{BT-sol-1})-(\ref{sol-3}) to the
equation (\ref{v-def}) yields the form of the potential:
\begin{equation}\label{pot-eta}
V(\eta)=\frac{V_0}{2}\int du \left(1+\zeta u^2\right)^4u^m,
\end{equation}
where
\begin{equation}\label{m-V0}
 m\equiv\frac{1}{3}(4\alpha+5\beta-15),\hspace{12mm}V_0\equiv-\frac{(1+\omega)(\alpha+2\beta)\beta^2\phi_0}{12b_0^2},
\end{equation}
\begin{equation}\label{u}
u(\eta)\equiv\left\{
 \begin{array}{c}
 tanh\left(\frac{\eta}{2}\right)
\hspace{12mm} {\rm for}\hspace{5mm} \zeta=-1,\\\\
tan\left(\frac{\eta}{2}\right)
\hspace{12mm} {\rm for}\hspace{5mm} \zeta=+1.
 \end{array}\right.
\end{equation}

Integrating the right hand side of (\ref{pot-eta}) yields
\begin{eqnarray}\label{pot-eta-2}
V(\eta)\!&=&\!V_0
u^m\sum^4_{n=0}\left\{\zeta^n\dbinom{4}{n}
\frac{u^{2n+1}}{m+(2n+1)}\right\},
 \end{eqnarray}
 where, without loss of generality, we have set the integration constant equal to zero.

 We should note that, in some particular cases, $V(\eta)$ vanishes:
 (i) $\alpha=-2\beta$, in which $\psi(\eta)$ takes constant values
 and therefore all the solutions for $\zeta=\pm1$ in the previous section as well as this
 section reduce to their counterparts obtained in the context of the standard BD theory in $4D$ space-time;
 (ii) $\omega=-1$: for this particular case, there are similarities between
 the scalar-tensor theories and supergravity~\cite{Faraoni.book}
 (it has been believed that, for this particular value of the BD coupling parameter, the standard BD theory
  can be considered as a low energy limit of the bosonic string
  theory~\cite{Faraoni.book,BD12}); and (iii) $\beta=0$: in this case,
 the BD scalar field takes constant values, therefore, our solutions in the
 previous section reduce to the corresponding Kantowski-Sachs and LRS Bianchi type III
 cosmological models in a $5D$ space-time obtained in general relativity, and
 consequently, the solutions of the present section describe the behavior
  of the quantities for the Kantowski-Sachs and LRS Bianchi type III
  cosmological models in the context of the induced-matter theory.
  We will further investigate the case (iii) in this paper.

Finally, substituting exact solutions
(\ref{BT-sol-1})-(\ref{sol-3}) and the induced scalar potential (\ref{pot-eta-2}) to
relations (\ref{t-00-eta-general})-(\ref{t-22-eta-general}), we obtain the
components of the induced matter in terms of $\eta$
\begin{eqnarray}\label{t-00-eta}
\rho(\eta)&=&
\Big[(\beta+2)+(\beta-2)\zeta u^2\Big]T_0u^{m}\left(1+\zeta u^2\right)^3\cr
\cr
&+&T_0(1+\omega)\beta^2u^{m}\sum^4_{n=0}\zeta^n\dbinom{4}{n}
\frac{u^{2n+1}}{m+(2n+1)},
\\\nonumber
\\\nonumber
\\
\label{t-11}
P_1(\eta)\!\!&=&\!\!T_0\!\!\left\{\left[\frac{2}{3}(2\alpha+\beta)\right]u\left(1+\zeta u^2\right)^4
-(1+\omega)\beta^2\sum^4_{n=0}\zeta^n\dbinom{4}{n}
\frac{u^{2n+1}}{m+(2n+1)}\right\}u^{m},
\\\nonumber
\\\nonumber
\\
\label{t-22}
P_2(\eta)&=&\frac{T_0}{3}\left[(3-2\alpha-\beta)-(3+2\alpha+\beta)\zeta u^2\right]u^{m+1}
\left(1+\zeta u^2\right)^3\cr
\cr
&-&T_0(1+\omega)\beta^2u^{m}\sum^4_{n=0}\zeta^n\dbinom{4}{n}
\frac{u^{2n+1}}{m+(2n+1)},
\end{eqnarray}
where
\begin{eqnarray}\label{T0}
T_0\equiv\frac{\phi_0(\alpha+2\beta)}{192\pi b_0^2}.
\end{eqnarray}
 In general, we see that the different
 components of $T^{k[{\rm BD}]}_{\,\,\,k}$ (where $k=1,2$, with no sum) are not equal, therefore,
the induced matter cannot be considered as a perfect fluid.

In the MBDT, the induced energy momentum tensor should also be conserved.
Its conservation for three cases, in terms of the cosmic time, can be written as
\begin{eqnarray}
\label{EMT-Cons-1}
\dot{\rho}+\sum^3_{i=1}\left(\rho+P_i\right)H_i=0,
\end{eqnarray}
where $H_1=\dot{a}/a$ and $H_2=H_3=\dot{b}/b$ are the directional Hubble parameters.
From (\ref{conformal}), (\ref{BT-sol-1})-(\ref{sol-3}), it is easy to show
that (\ref{EMT-Cons-1}), in terms of the new time coordinate, is given by
\begin{eqnarray}
\label{EMT-Cons-2}
u\rho'(\eta)+\frac{1}{3}\left(2\alpha +\beta\right)\left(1+\zeta u^2\right)\Big[P_1(\eta)-P_2(\eta)\Big]
+\left(1-\zeta u^2\right)\Big[\rho(\eta)+P_2(\eta)\Big]
=0,
\end{eqnarray}
where $u(\eta)$ is given by (\ref{u}).
Substituting (\ref{t-00-eta})-(\ref{t-22}) to equation (\ref{EMT-Cons-2}) and then
using (\ref{sol-5}), it is straightforward to show that the
above equality is satisfied for both Kantowski-Sachs and LRS Bianchi type III models.

 It is worthwhile to investigate the properties of a few
  physical quantities, which are important in observational
cosmology.
For instance, let us define the spatial volume ${\rm V}_s$, average
  Hubble parameter ${\rm H}$, mean anisotropy parameter ${\rm A}_h$, the deceleration
parameter $q$ and the expansions for scalar
expansion $\theta$ and the shear scalar $\sigma^2$, in terms
of the cosmic time, as
\begin{eqnarray}\nonumber
V_s\!\!&=&\!\!A^3(t)=a(t)b^2(t), \hspace{5mm}\theta=3H=\left(\frac{\dot{a}}{a}+\frac{2\dot{b}}{b}\right),\\\nonumber
A_h\!\!&=&\!\!\frac{1}{3}\sum^3_{i=1}\left(\frac{\Delta H_i}{H}\right)^2, \hspace{5mm} {\rm where}\hspace{5mm}\Delta H_i=H_i-H,\\\nonumber
q\!\!&=&\!\!\frac{d}{dt}\left(\frac{1}{H}\right)-1=-\frac{A\ddot{A}}{\dot{A}^2},\\
\label{phys.quant}
\sigma^2\!\!&=&\!\!\frac{1}{2}\sigma_{ij}\sigma^{ij}=\frac{1}{3}\left(H_1^2+H_2^2-2H_1H_2\right),
\end{eqnarray}
where $i,j=1,2,3$ and ${\rm A(t)}$ is mean scale factor of the universe.

Substituting solutions (\ref{BT-sol-1})-(\ref{sol-3}) in
the corresponding relations (\ref{phys.quant}), we obtain
the above quantities in terms of $\eta$:
\begin{eqnarray}\nonumber
V_s(\eta)\!\!&=&\!\!A^3(\eta)=a_0b_0^2\left(\frac{2u}{1+\zeta u^2}\right)^2,\\\nonumber
\\\nonumber
\theta(\eta)\!\!&=&\!\!3H(\eta)=\left(\frac{1-u^4}{2b_0 u^2}\right)u^{\frac{1}{3}(2\alpha+\beta)},\\\nonumber
\\\nonumber
A_h(\eta)\!\!&=&\!\!(2\alpha+\beta)\left(\frac{1+\zeta u^2}{1-\zeta u^2}\right)
\left[\left(\frac{2\alpha+\beta}{2}\right)\left(\frac{1+\zeta u^2}{1-\zeta u^2}\right)-1\right]+\frac{1}{2},\\\nonumber
\\\nonumber
q(\eta)\!\!&=&\!\!\frac{(4+2\alpha+\beta)u^4+4(1+\zeta u^2)-(2\alpha+\beta)}{2(1-\zeta u^2)^2},\\\nonumber
\\\label{phys.quant-2}
\sigma^2(\eta)\!\!&=&\!\!\frac{1}{3b_0^2}\left(\frac{1+\zeta u^2}{2u}\right)^4
\left[(2\alpha+\beta)-\left(\frac{1-\zeta u^2}{1+\zeta u^2}\right)\right]^2u^{\frac{2}{3}(2\alpha+\beta)}.
\end{eqnarray}

As mentioned, when the BD coupling parameter goes to infinity, the BD solutions may reduce to the corresponding
 counterparts in general relativity.
Let us close this subsection by studying this particular case.
From (\ref{omega}), we see that when $\beta$ goes
 to zero, then $\vert\omega\vert$ tends to infinity, and consequently, the BD
scalar field takes constant values. In this limit, from~(\ref{sol-5}), we get
\begin{equation}\label{101}
\alpha=\pm\frac{\sqrt{3}}{2}.
 \end{equation}
Therefore, letting $\beta=0$ and substituting the values of $\alpha$ from the above
relation to the solutions~(\ref{BT-sol-3}), (\ref{sol-1}) and (\ref{sol-3}), we obtain
\begin{itemize}
  \item {$\zeta=-1$:}\\
  \begin{eqnarray}
\label{IMT-sol-1}
a(\eta)=a_0u^{\pm\frac{2\sqrt{3}}{3}},\hspace{5mm}
b(\eta)=b_0sinh\eta\,\, u^{\mp\frac{\sqrt{3}}{3}},\hspace{5mm}
\psi(\eta)=\psi_0\,u^{\mp\frac{\sqrt{3}}{3}},
\end{eqnarray}
\item {$\zeta=+1$:}
  \begin{eqnarray}
\label{IMT-sol-2}
a(\eta)=a_0u^{\pm\frac{2\sqrt{3}}{3}},\hspace{5mm}
b(\eta)=b_0sin\eta \,\,u^{\mp\frac{\sqrt{3}}{3}},\hspace{5mm}
\psi(\eta)=\psi_0u^{\mp\frac{\sqrt{3}}{3}},
\end{eqnarray}
\end{itemize}
where $u=u(\eta)$ was introduced by relations (\ref{u}).
It is straightforward to show that solutions (\ref{IMT-sol-1}) and (\ref{IMT-sol-2}) satisfy
equations (\ref{prime-2})-(\ref{prime-5}). Moreover, for this particular
case, assuming $c_1=0$, (\ref{prime-1}) yields an identity, $0=0$.

In addition, from (\ref{pot-eta}), we find that the induced scalar potential
vanishes, and therefore, from (\ref{t-00-eta})-(\ref{t-22}), the
components of the induced energy momentum tensor reduce to
\begin{eqnarray}\label{t-00-IMT}
\rho(\eta)\!&=&\!2 T_0\left(1-\zeta u^2\right)\left(1+\zeta u^2\right)^3
u^{m+1},
\\\nonumber
\\\nonumber
\\
\label{t-11-IMT}
P_1(\eta)\!&=&\!
\pm\frac{2\sqrt{3}}{3}T_0\left(1+\zeta u^2\right)^4u^{m+1},\\\nonumber
\\\nonumber
\\
\label{t-22-IMT}
P_2(\eta)\!&=&\!
\frac{T_0}{3}\left[(3\mp\sqrt{3})-3(3\pm\sqrt{3})\zeta u^2\right]
\left(1+\zeta u^2\right)^3u^{m+1},
\end{eqnarray}
where
\begin{eqnarray}\label{T0}
T_0\equiv\frac{\pm \sqrt{3}\phi_0}{384\pi b_0^2},
\hspace{10mm}m=\frac{1}{3}(\pm 2\sqrt{3}-15).
\end{eqnarray}

\subsection{Effective cosmologies for $\zeta=0$}

In what follows, we obtain the corresponding induced scalar potential, the effective
matter and the other physical quantities.
Using relations (\ref{sol-3-BT1}), equation (\ref{pot-eta-all}) can be written as
\begin{equation}\label{pot-eta-BT1}
\frac{dV(\phi)}{d\phi}=-\frac{4\beta(1+\omega)(\alpha+2\beta-\gamma-3)}{3b_0^2}
\left(\frac{\phi}{\phi_0}\right)^{\frac{2\xi}{3\beta}},
\end{equation}
 where, for convenience, we introduced a new parameter
\begin{equation}
\label{xi}
\xi\equiv2\alpha+\beta+\gamma-3.
\end{equation}
Setting the integration constant equal to zero, (\ref{pot-eta-BT1}) gives
\begin{equation}\label{V-BT1}
V(\phi)=\left\{
 \begin{array}{c}
-\frac{4\phi_0\beta^2(1+\omega)(\alpha+2\beta-\gamma-3)}{b_0^2(2\xi+3\beta)}
\left(\frac{\phi}{\phi_0}\right)^{\frac{2\xi}{3\beta}}
\hspace{12mm} {\rm for}\hspace{10mm} 2\xi+3\beta\neq0,\\\\
-\frac{4\phi_0\beta(1+\omega)(\alpha+2\beta-\gamma-3)}{3b_0^2}{\rm ln}\phi
\hspace{22mm} {\rm for}\hspace{10mm} 2\xi+3\beta=0.
 \end{array}\right.
\end{equation}
In addition, by substituting the BD scalar field from (\ref{sol-3-BT1}) to the above
relation, the induced scalar potential is written in terms of $\eta$
\begin{equation}\label{V-BT1-2}
V(\eta)=\left\{
 \begin{array}{c}
-\frac{4\phi_0\beta^2(1+\omega)(\alpha+2\beta-\gamma-3)}{b_0^2(2\xi+3\beta)}
\eta^{\frac{1}{3}(2\xi+3\beta)}
\hspace{12mm} {\rm for}\hspace{10mm} 2\xi+3\beta\neq0,\\\\
-\frac{4\phi_0\beta(1+\omega)(\alpha+2\beta-\gamma-3)}{3b_0^2}{\rm ln}(\phi_0\eta^\beta)
\hspace{15mm} {\rm for}\hspace{10mm} 2\xi+3\beta=0.
 \end{array}\right.
\end{equation}
As the logarithmic branch of the induced scalar potential leads to an effective matter which
 is complicated to discuss regarding the energy conditions,
let us just investigate the power law solutions.
However, for the power-law branch, substituting relations (\ref{sol-3-BT1})
and (\ref{V-BT1-2}) to (\ref{t-00-eta-general})-(\ref{t-22-eta-general}), the
components of the induced energy momentum tensor for $\zeta=0$ are given by
\begin{eqnarray}\label{t-00-eta-BT1}
\rho(\eta)\!&=&\!\frac{\phi_0(\alpha+2\beta-\gamma-3)}{4\pi b_0^2}
\left[\frac{\beta-\gamma-1}{3}+\frac{\beta^2(1+\omega)}{2\xi+3\beta}\right]\eta^{\frac{1}{3}(2\xi+3\beta)},
\\\nonumber
\\\nonumber
\\
\label{t-11-BT1}
P_1(\eta)\!&=&\!
\frac{\phi_0(\alpha+2\beta-\gamma-3)}{4\pi b_0^2}
\left[\frac{2\xi+3(1-\gamma)}{9}-\frac{\beta^2(1+\omega)}
{2\xi+3\beta}\right]\eta^{\frac{1}{3}(2\xi+3\beta)},\\\nonumber
\\\nonumber
\\
\label{t-22-BT1}
P_2(\eta)\!&=&\!
-\frac{\phi_0(\alpha+2\beta-\gamma-3)}{4\pi b_0^2}
\left[\frac{\xi+3}{9}+\frac{\beta^2(1+\omega)}
{2\xi+3\beta}\right]\eta^{\frac{1}{3}(2\xi+3\beta)}.
\end{eqnarray}
In order to check the conservation of the
induced energy momentum tensor, let us proceed as follows: using (\ref{sol-3-BT1}) and the above
relations, equation (\ref{EMT-Cons-1}) is rewritten in terms of the new time coordinate as
\begin{eqnarray}
\label{EMT-Cons-2-BT1}
\eta\rho'(\eta)-(\gamma+1)\Big[\rho(\eta)+P_1(\eta)\Big]+\frac{2}{3}\left(\xi+3\right)\Big[P_1(\eta)-P_2(\eta)\Big]=0.
\end{eqnarray}
From (\ref{t-00-eta-BT1})-(\ref{t-22-BT1}) and the constraint
(\ref{con.BT1}), it is easy to show that (\ref{EMT-Cons-2-BT1}) is satisfied.

For $\zeta=0$, it is straightforward to show that the physical
quantities (\ref{phys.quant}), in terms of the new time coordinate are given by
\begin{eqnarray}\nonumber
V_s(\eta)\!\!&=&\!\!a_0b_0^2\eta^{-(1+\gamma)},\\\nonumber
\theta(\eta)\!\!&=&\!\!3H(\eta)=-\frac{\gamma+1}{b_0}\eta^{\frac{\xi}{3}},\\\nonumber
A_h(\eta)\!\!&=&\!\!2\left(\frac{2\alpha+\beta-1}{1+\gamma}\right)^2,\\\nonumber\\\nonumber
q(\eta)\!\!&=&\!\!\frac{2\alpha+\beta-4}{\gamma+1},\\\label{phys.quant-BT1}
\sigma^2(\eta)\!\!&=&\!\!\frac{(2\alpha+\beta-1)^2}{3b_0^2}\eta^{\frac{2\xi}{3}}.
\end{eqnarray}


\section{Analytic solutions and cosmic time}

\label{Deparametrized}
\begin{figure}
\centering\includegraphics[width=2.6in]{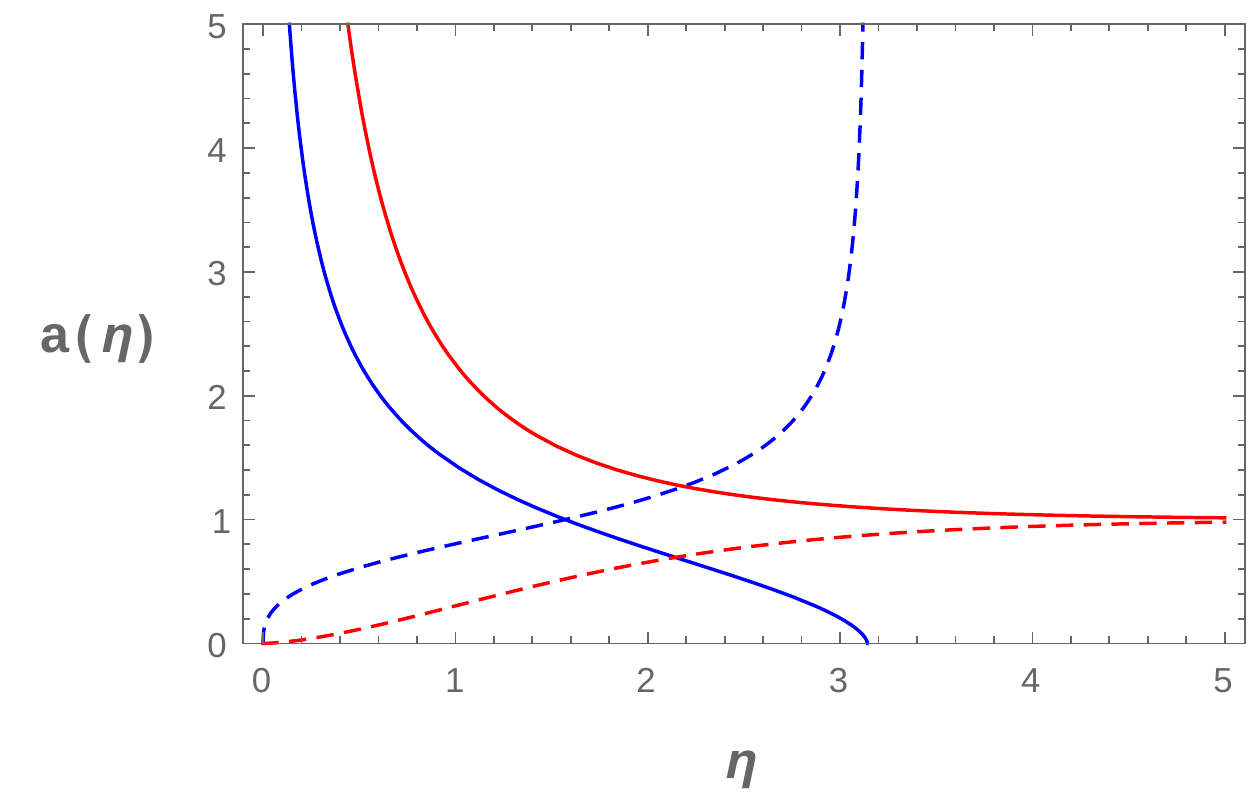}
\hspace{5mm}
\centering\includegraphics[width=2.6in]{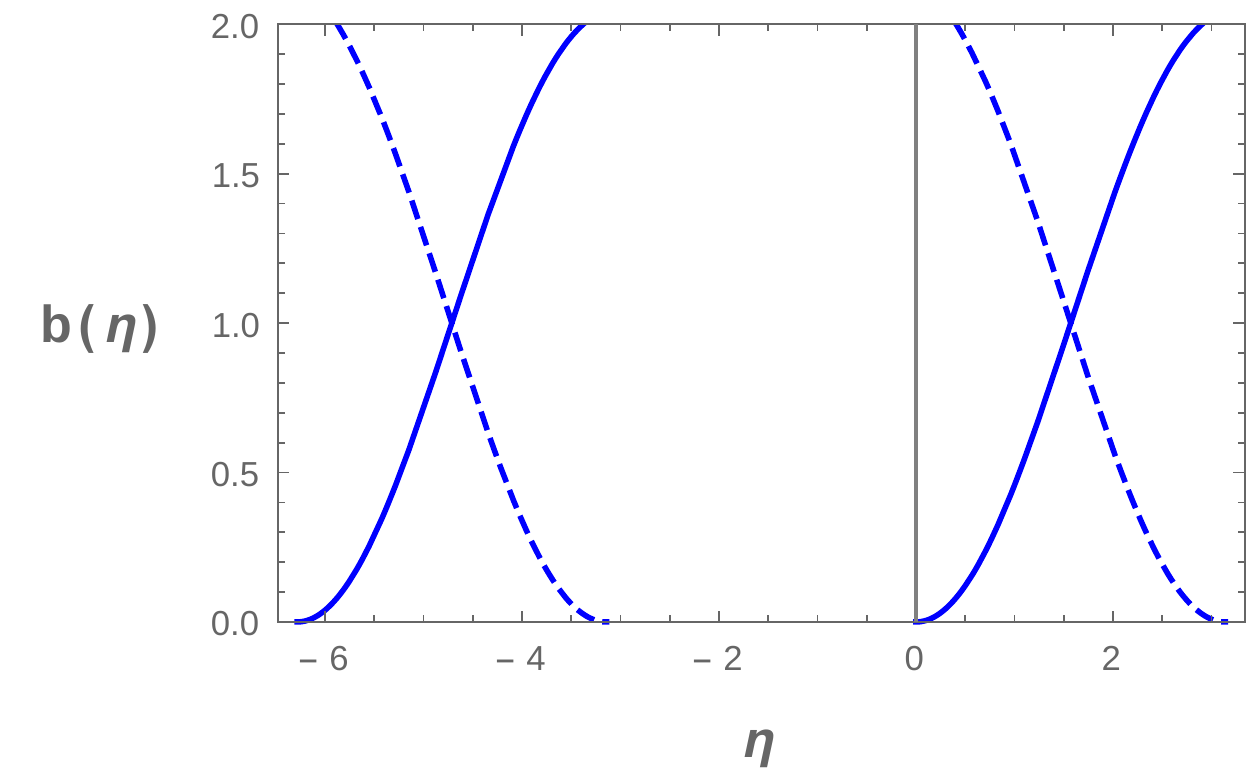}
\centering\includegraphics[width=2.6in]{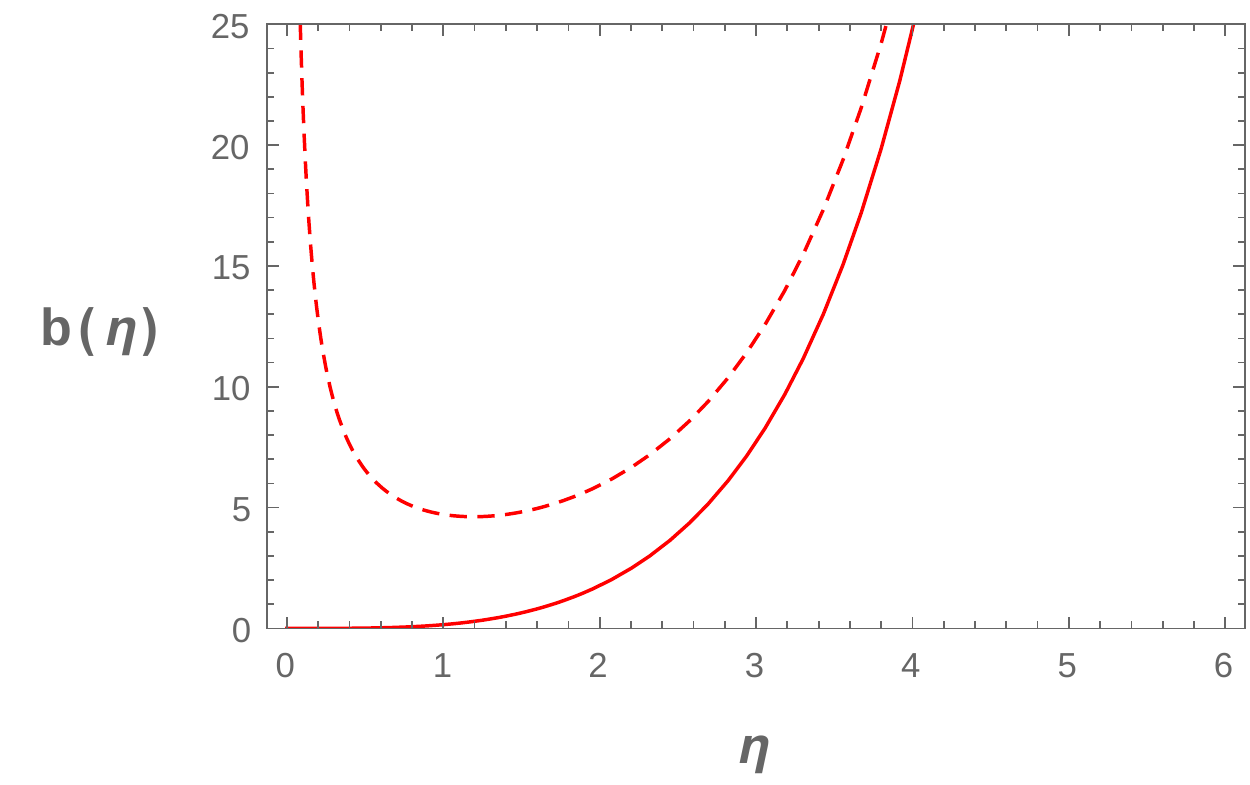}
\caption{The behavior
of the scale factors $a$ and $b$ associated with $\zeta=-1$ (red curves) and $\zeta=+1$ (blue curves)
in terms of the parametric time $\eta$ (which is related to the cosmic time by equation (\ref{conformal})).
We have set $a_0=1=b_0$, and the dashed and solid
curves are for $2\alpha+\beta>0$ and $2\alpha+\beta<0$, respectively.}
\label{a-b-BT3-KS}
\end{figure}

As seen from (\ref{BT-sol-1}) and (\ref{sol-1}), for the
cases $\zeta=\pm1$, $b(\eta)$ is a complicated function of $\eta$, and thus
finding analytical solutions in terms of the cosmic time, $t$, requires calculating complicated
integrals, which even for the special cases are almost impossible.
Hence, we shall restrict ourselves to depict the behaviour of quantities
in terms of the parametric time, which is related to the cosmic time by
equation (\ref{conformal}). For instance, figure \ref{a-b-BT3-KS} shows the
 behavior of the scale factors $a$ and $b$ in terms of $\eta$.
 Moreover, we should note that the plot of $a(\eta)$ also depicts
 the behavior of $\phi$ and $\psi$ (in terms of $\eta$) by setting $2(2\alpha+\beta)/3\rightarrow \beta$
 and $2(2\alpha+\beta)/3\rightarrow -2(\alpha+2\beta)/3$, respectively.
 Furthermore, figure \ref{a-b-BT3-KS} also implies
 that the behaviors of the scale factors (in terms of $\eta$) can
 be stated in terms of only two parameters; for instance, herein, we took
  parameters $\alpha$ and $\beta$, and remove the BD coupling constant, because
  the latter is constrained by relation (\ref{omega}).

However, for the solutions associated with the LRS Bianchi type I, the
de-parametrization procedure is feasible and we will show that our
solutions can be considered as extended versions of those obtained, by
assuming specific ansatzes, in the context of either the
standard BD theory or even the MBDT, see for instance~\cite{RFS11}.
In what follows, let us investigate the case $\zeta=0$.
Substituting $b(\eta)$ from (\ref{sol-3-BT1}) to (\ref{conformal}) and then
integrating both sides of it, we obtain
\begin{equation}\label{t-BT1}
t(\eta)=\left\{
 \begin{array}{c}
-\frac{3b_0}{\xi}\left(\eta^{-\frac{\xi}{3}}-\eta_0^{-\frac{\xi}{3}}\right)
\hspace{12mm} {\rm for}\hspace{10mm} \xi\neq0,\\\\
b_0\rm{ln(\frac{\eta}{\eta_0})}\hspace{33mm} {\rm for}\hspace{10mm} \xi=0,
 \end{array}\right.
\end{equation}
where $\eta_0$ is an integration constant. Without loss of generality, we can set the
integration constant $\eta_0$ equal to zero.
We also should note that as we have not been investigating the logarithmic
induced scalar potential, thus, according to (\ref{V-BT1-2}), the
constraint $4\alpha+5\beta+2\gamma-6=2\xi+3\beta\neq0$ must
hold for all solutions.

Therefore, from relations (\ref{t-BT1}), we get all the solutions in
terms of the cosmic time. In the following subsections, we will
show that there are two types of solutions, namely, the power-law and exponential-law solutions.
Subsequently, we will investigate the physical properties of each case.

Let us introduce new parameters. Concretely,
\begin{eqnarray}
\label{BT1-power-law-eos}
w_i=\frac{P_i(t)}{\rho(t)},
\end{eqnarray}
where $\rho(t)$ and $P_i(t)$ are the components of the induced
energy momentum tensor in terms of the cosmic time and $w_i$ are
the directional equation of state parameters along the axes.
  Moreover, we set $w$ as the deviation-free equation of state parameter
 associated with the induced matter. In order to parameterize the deviation from the
 isotropy, we set $w=w_1$ and then we introduce the skewness parameters
 as $\delta_j=w_j-w$ (where $j=2,3$), which indicate the deviation from $w$ along the other two directions.
 As in our model $P_2=P_3$, we therefore obtain $\delta_2=\delta_3\equiv\delta $.

Using relations (\ref{sol-3-BT1}) and (\ref{t-BT1}), we obtain
power-law and exponential-law solutions in terms of the cosmic time. Let us
investigate them in separated subsections.

\subsection{Power law solutions ($\xi\neq0$)}

The solutions which are power-law forms of the cosmic time are given by (applying the
specific assumptions, such as {\it effective pressure} and the particular law
of variation of the Hubble parameter, in Ref. \cite{RFS11} yielded
 more different and more restricted solutions
than those we will investigate in this paper)
\begin{eqnarray}
\label{BT1-power-law-a}
a(t)&=&a_0\left(-\frac{\xi t}{3b_0}\right)^{\frac{-2\xi+3(\gamma-1)}{\xi}},\hspace{7mm}
b(t)=\left(-\frac{\xi t}{3b_0 }\right)^{\frac{\xi+3}{\xi}},\\\nonumber
\\
\label{BT1-power-law-phi}
\phi(t)&=&\phi_0\left(-\frac{\xi t}{3b_0 }\right)^{-\frac{3\beta}{\xi}},\hspace{16mm}
\psi(t)=\psi_0\left(-\frac{\xi t}{3b_0 }\right)^{\frac{\xi+3(\beta-\gamma-1)}{\xi}}.
\end{eqnarray}

Moreover, by using (\ref{t-BT1}) in relations (\ref{V-BT1-2})-(\ref{t-22-BT1}), it is easy to show that the
induced scalar potential, the effective energy density, the (deviation-free) equation of state and skewness parameters
associated with the power-law solutions are written as
\begin{eqnarray}
\label{V-BT1-dep}
V(t)\!&=&\!V_p
\left(-\frac{\xi t}{3b_0 }\right)^{-\frac{2\xi+3\beta}{\xi}},\hspace{5mm} V_p\equiv-\frac{4\phi_0\beta^2(1+\omega)(\alpha+2\beta-\gamma-3)}{b_0^2(2\xi+3\beta)},
\\\nonumber
\\\nonumber
\\
\label{t-00-eta-BT1-dep}
\rho(t)\!&=&\!\rho_p
\left(-\frac{\xi t}{3b_0 }\right)^{-\frac{2\xi+3\beta}{\xi}},\hspace{5mm}\rho_p\equiv\frac{\phi_0(\alpha+2\beta-\gamma-3)}{4\pi b_0^2}
\left[\frac{\beta-\gamma-1}{3}+\frac{\beta^2(1+\omega)}{2\xi+3\beta}\right],
\\\nonumber
\\\nonumber
\\
\label{w-BT1-dep}
w_p\!&=&\!\frac{1}{3}\left[\frac{16\alpha^2+4\alpha(7\beta+\gamma-9)+\beta^2(1-9\omega)-\beta(\gamma+27)-2\gamma^2+18}
{4\alpha(\beta-\gamma-1)+\beta^2(3\omega+8)-\beta(3\gamma+11)-2(\gamma-3)(\gamma+1)}\right],
\\\nonumber
\\\nonumber
\\
\label{delta-BT1-dep}
\delta_p\!&=&\!-\frac{(2\alpha+\beta-1)(2\xi+3\beta)}{4\alpha(\beta-\gamma-1)
+\beta^2(3\omega+8)-\beta(3\gamma+11)-2(\gamma-3)(\gamma +1)},
\end{eqnarray}
where $w_p$ and $\delta_p$ are the (deviation-free) equation of state and skewness
parameters associated with this power-law solution.
The solutions (\ref{BT1-power-law-a})-(\ref{delta-BT1-dep})
constitute a new Bianchi type I cosmological dynamics in MBDT.

Relations (\ref{BT1-power-law-a}), (\ref{t-00-eta-BT1-dep})-(\ref{delta-BT1-dep}) and
the constraint (\ref{omega.BT1}) allow to extract the energy momentum tensor conservation law (\ref{EMT-Cons-1}), identically.

Furthermore, re-employing relations (\ref{t-BT1}), the other
physical quantities, in terms of the cosmic time, are given by
\begin{eqnarray}\nonumber
V_s(t)\!\!&=&\!\!A(t)^3=a_0b_0^2\left(-\frac{\xi t}{3b_0 }\right)^{\frac{3(1+\gamma)}{\xi}},\hspace{10mm}
\theta(t)=3H(t)=\frac{3(\gamma+1)}{\xi t},\\\nonumber
\\\nonumber
A_h\!\!&=&\!\!2\left[\frac{\xi+3-(\gamma+1)}{\gamma+1}\right]^2,\hspace{22mm}
q=\frac{\xi-(\gamma+1)}{\gamma+1},\\\nonumber\\\label{phys.quant-BT1-dep}
\sigma^2\!\!&=&\!\!3\left[\frac{\xi+3-(\gamma+1)}{\xi t}\right]^2.
\end{eqnarray}
Let us explain the time behaviors of the physical quantities (by assuming $a_0>0$):
(i) we observe that for $t=0$, there is a singularity. For any arbitrary
value of $b_0$, if $\xi(1+\gamma)>0$, the spatial volume expands, while for $\xi(1+\gamma)<0$, it always contracts;
(ii) the Hubble parameter goes to zero when $t\rightarrow\infty$;
(iii) the shear and expansion scalars diverge at $t=0$ and they vanish
when $t\rightarrow\infty$; (iv) from the relation associated with the
deceleration parameter, we see that for $\frac{\xi}{\gamma+1}>1$, the mean scale factor of
the universe decelerates, while for  $0<\frac{\xi}{\gamma+1}<1$, we get an accelerating mean
scale factor, which is in accordance with the observational data \cite{Riess98, WMAP13, Planck15};
(v) from relations (\ref{phys.quant-BT1-dep}), it is seen that
$\frac{\sigma^2}{\theta^2}=\frac{1}{3}\frac{\xi+3}{\gamma+1}-1={\rm constant}$, which indicates
that the model does not approach isotropy when the cosmic time takes large values.

In the rest of this subsection, let us investigate the reduced
isotropic cosmological model resulted from the power-law solutions.

 We first remind that for all the solutions of this
 section, we have a constraint as $4\alpha+5\beta+2\gamma-6=2\xi+3\beta\neq0$.
 In order to get an isotropic fluid, we should set $\delta_p=0$, and from
 relation (\ref{delta-BT1-dep}), we obtain either $\beta=-\frac{2}{5}(2\alpha +\gamma -3)$
 or $\beta=1-2\alpha$. However, the former value is not acceptable because it is in contradiction with the mentioned constraint.
 For the latter value of $\beta$, from relations (\ref{BT1-power-law-a})-(\ref{phys.quant-BT1-dep}), we get
 $A_h=0=\sigma^2$, as expected. The set of resulted solutions are summarized as
 \begin{eqnarray}
 \label{iso-power-law-1}
ds^2&=&-dt^2+\left[\left(\frac{2-\gamma}{3b_0}\right)t\right]^{\frac{2(\gamma+1)}{\gamma-2}}\left(dr^2+d\Omega^2_{\zeta=0}\right),\\\nonumber
\\
\label{iso-power-law-phi}
\phi(t)\!&=&\!\phi_0\left[\left(\frac{2-\gamma}{3b_0}\right)t\right]^{\frac{6\alpha-3}{\gamma-2}},\hspace{15mm}
\psi(t)=\psi_0\left[\left(\frac{2-\gamma}{3b_0}\right)t\right]^{-\frac{2(3\alpha+\gamma+1)}{\gamma-2}},\\\nonumber
\\
\rho\!&=&\!\rho_p\left[\left(\frac{2-\gamma}{3b_0}\right)t\right]^{\frac{6\alpha-2\gamma+1}{\gamma-2}},\hspace{10mm}
\rho_p\!=\!-\frac{\phi_0(1+\gamma)(3\alpha+\gamma+1)(6\alpha+4\gamma+1)}{12\pi b_0^2(6\alpha-2\gamma+1)}, \\\nonumber
\\
\label{iso-power-law-w}
w_p\!&=&\!-\frac{6\alpha+\gamma+4}{3(\gamma+1)},\hspace{28mm} q=-\frac{3}{1+\gamma},\\\nonumber
\\
\label{iso-power-law-omega}
\omega\!&=&\!-\frac{2\left[12\alpha^2+(4\gamma-2)\alpha+\gamma^2+3\gamma+2\right]}{3(1-2\alpha)^2},
\end{eqnarray}
where we have set $a_0=1$. It is seen that relation (\ref{iso-power-law-w}) is an equation of state of a barotropic fluid.

 In a particular case where $\beta=0$, then, $\alpha=1/2$, and therefore the BD scalar field takes a constant
value and the BD coupling parameter goes to infinity.
In this case, in order to find the exact solutions, we should start
 from the field equations (\ref{dot-1})-(\ref{dot-5}). It is straightforward to show that they are satisfied
 whether $\gamma=-1$ or $\gamma=-4$. The former case leads to a
 static universe which is not of interest in this paper. For $\gamma=-4$, we obtain the unique solution
 associated with the general relativistic field equations for a $5D$
 spatially flat FRW universe in vacuum
\begin{eqnarray}\nonumber
ds^2&=&-dt^2+\left(\frac{2t}{b_0}\right)\left(dr^2+d\Omega^2_{\zeta=0}\right),\\
\psi(t)&=&\psi_0\left(\frac{2t}{b_0}\right)^{-\frac{1}{2}},\hspace{20mm}
\phi(t)=\phi_0={\rm constant}.\label{iso-power-law-IMT-1}
\end{eqnarray}
In this case, we get a decelerating scale factor for the universe. Moreover, the fifth
dimension contracts when the cosmic time increases.
Inserting the above relations in (\ref{t-00}) and (\ref{t-ii}) (the induced scalar
potential, without loss of generality, can be assumed equal to zero), we retrieve
\begin{eqnarray}
\label{iso-power-law-IMT-2}
\rho=\frac{3\phi_0}{32\pi}\frac{1}{t^2},\hspace{19mm}
w_p=\frac{1}{3},
\end{eqnarray}
which corresponds to a radiative fluid. These results associated with the power-law
solutions for $\beta=0$ are similar to those obtained in \cite{RFM14}.
Moreover, other cases associated with a spatially flat FRW universe, in which
the effective matter can play the role of an extended quintessence, radiation
and dust, have been widely investigated in MBDT \cite{RFM14,RM16}.

As another particular case of the isotropic solutions
(\ref{iso-power-law-1})-(\ref{iso-power-law-omega}), we investigate the universe in which the stiff fluid is dominant.
To the best of our knowledge this case has not been studied in the context of MBDT.
 By setting $w_p=1$ in (\ref{iso-power-law-w}), we obtain a relation for $\alpha$
 in terms of $\gamma$ as $\alpha=-(4\gamma+7)/6$ and consequently,
 $\beta$ can also be written in terms of $\gamma$ as $\beta=2(2\gamma+5)/3$.
 Substituting these values of
$\alpha$ and $\beta$ in the relations associated with the isotropic
fluid, i.e., (\ref{iso-power-law-1})-(\ref{iso-power-law-omega}), all the
quantities can be written in terms of $\gamma$

 \begin{eqnarray}
 \label{iso-power-law-stiff-1}
ds^2&=&-dt^2+\left[\left(\frac{2-\gamma}{3b_0}\right)t\right]^{\frac{2(\gamma+1)}{\gamma-2}}\left(dr^2+d\Omega^2_{\zeta=0}\right),\\\nonumber
\\
\label{iso-power-law-stiff-2}
\phi(t)&=&\phi_0\left[\left(\frac{2-\gamma}{3b_0}\right)t\right]^{-\frac{2(2\gamma+5)}{\gamma-2}},\hspace{20mm}
\psi(t)=\psi_0\left[\left(\frac{2-\gamma}{3b_0}\right)t\right]^{\frac{2\gamma+5}{\gamma-2}},\\\nonumber
\\
\rho&=&\frac{\phi_0(2\gamma+5)}{24\pi b_0^2}\left[\left(\frac{2-\gamma}{3b_0}\right)t\right]^{\frac{6(\gamma+1)}{2-\gamma}},\\\nonumber
\\
\label{iso-power-law-stiff-3}
\omega&=&-\frac{11\gamma^2+55\gamma+62}{2(2\gamma+5)^2},\hspace{35mm} q=-\frac{3}{1+\gamma}.
\end{eqnarray}

Let us now study three important particular cases, concerning the stiff fluid:
\begin{itemize}
  \item  When $\gamma=-5/2$ then $\omega$ goes to infinity, $\alpha=1/2$, $\beta=0$ and
the BD scalar field takes constant values. However, as we discussed
above equation (\ref{iso-power-law-IMT-1}), with such values of the parameters $\alpha$, $\beta$ and
$\gamma$, equations (\ref{dot-1})-(\ref{dot-5}) are not satisfied.
This result can be as an obvious evidence to suggest that general relativity is not always
recovered from the BD theory, when the BD coupling
parameter goes to infinity~\cite{BR93,BS97,Faraoni99}.

  \item With $\omega=-1$,
  from (\ref{iso-power-law-stiff-3}) we
get either $\gamma=-1$ or $\gamma=-4$. The former leads to a static universe
 which is not of interest in this work. However, for the latter, we obtain the following solutions
\begin{eqnarray}
 \label{power-law-stiff-omega1}
ds^2&=&-dt^2+\left(\frac{2t}{b_0}\right)\left(dr^2+d\Omega^2_{\zeta=0}\right),
\hspace{15mm}\psi(t)=\psi_0\left(\frac{2t}{b_0}\right)^{\frac{1}{2}},\\\nonumber
\\
\label{power-law-stiff-omega2}
\phi(t)&=&\frac{\phi_0b_0}{2t},\hspace{35mm}
\hspace{20mm}
\rho(t)=-\frac{b_0\phi_0}{64\pi t^3},
\end{eqnarray}
and $q=1$, which describes a decelerating universe.

  \item  Letting $\omega=-4/3$, from (\ref{iso-power-law-stiff-3}), we
obtain either $\gamma=2$ (which is not acceptable) or $\gamma=-7$. The latter yields
\begin{eqnarray}
 \label{power-law-stiff-omega3}
ds^2&=&-dt^2+\left(\frac{3t}{b_0}\right)^{\frac{4}{3}}\left(dr^2+d\Omega^2_{\zeta=0}\right),
\hspace{15mm}\psi(t)=\left(\frac{3\psi_0t}{b_0}\right),\\\nonumber
\\
\label{power-law-stiff-omega4}
\phi(t)&=&\frac{\phi_0b_0^2}{9t^2},\hspace{35mm}
\hspace{20mm}\rho(t)=-\frac{\phi_0b_0^2}{216 \pi t^4},
\end{eqnarray}
and $q=3/4$, which, again, describes a decelerating universe.
\end{itemize}

  \subsection{ Exponential-law solutions ($\xi=0$)}
  \label{exp-iso}
  By focusing on the logarithmic branch in (\ref{t-BT1}), from relations
   (\ref{sol-3-BT1}), we get new solutions
  \begin{eqnarray}
\label{exp-a}
a(t)&=&a_e\,{\rm exp}\left[\left(\frac{1-\gamma}{b_0}\right)t\right],\hspace{27mm}
b(t)=b_e\,{\rm exp}\left(-\frac{t}{b_0}\right),\\
\label{exp-phi}
\phi(t)&=&\phi_e\,{\rm exp}\left(\frac{\beta t}{b_0}\right),\hspace{37mm}
\psi(t)=\psi_e\,{\rm exp}\left[\left(\frac{1+\gamma-\beta}{b_0}\right)t\right],
\end{eqnarray}
where $a_e\equiv a_0 \eta_0^{1-\gamma}$, $b_e\equiv b_0/\eta_0$, $\phi_e\equiv \phi_0 \eta_0^\beta$ and $\psi_e\equiv\psi_0\eta_0^{1-\beta+\gamma}$.
The induced potential
and the effective matter on a $4D$ hypersurface are

\begin{eqnarray}
\label{exp-V-BT1-dep}
V(t)\!&=&\!V_e\,\,{\rm exp}\left({\frac{\beta t}{b_0}}\right),
 \hspace{15mm} V_{e}\equiv-\frac{2\phi_0\eta_0^\beta(\omega +1)\beta(\beta-\gamma-1)}{b_0^2},
\\\nonumber
\\\nonumber
\\
\label{exp-t-00-eta-BT1-dep}
\rho(t)\!&=&\!
\rho_e\,\,{\rm exp}\left({\frac{\beta t}{b_0}}\right), \hspace{15mm} \rho_e\equiv \frac{\phi_0\eta_0^\beta(\beta-\gamma-1)\left[\beta(\omega+2)-(\gamma+1)\right]}{8\pi b_0^2},
\\\nonumber
\\\nonumber
\\
\label{exp-t-11-BT1-dep}
w_e\!&=&\!-\frac{\beta(\omega+1)+(\gamma-1)}{\beta(\omega+2)-(\gamma+1)},\\\nonumber
\\
\label{exp-t-11-BT1-delta}
\delta_e\!&=&\!\frac{\gamma-2}{\beta(\omega+2)-(\gamma+1)}.
\end{eqnarray}
Moreover, using the corresponding constraint associated with
this case ($\xi=0$) in (\ref{omega.BT1}), gives the following BD coupling
\begin{eqnarray}\label{omega-exp}
\omega=-\frac{2\left[\beta^2-\beta(\gamma+1)+\gamma^2+2\right]}{\beta^2}.
\end{eqnarray}
Applying relations (\ref{exp-a}), (\ref{exp-t-00-eta-BT1-dep}), (\ref{exp-t-11-BT1-dep}) and
the constraint (\ref{omega-exp}), it is straightforward to show
that the conservation law for the energy momentum tensor is satisfied identically.

Furthermore, relations (\ref{t-BT1}) provide
\begin{eqnarray}\nonumber
V_s(t)\!&=&\!A(t)^3=a_0b_0^2{\rm exp}\left[-\frac{(\gamma+1)t}{b_0}\right],\\\nonumber
\theta\!&=&\!3H(t)=-\frac{\gamma+1}{b_0},\hspace{25mm}A_h=2\left(\frac{2-\gamma}{\gamma+1}\right)^2,
\\
q\!&=&\!-1,\hspace{47mm}\sigma^2=3\left(\frac{2-\gamma}{3b_0}\right)^2.\label{phys.quant-BT1-dep-exp}
\end{eqnarray}
Let us start with focusing on the physical properties of the solutions when
the effective matter is an anisotropic fluid.
Subsequently, we will discuss on the cases where the induced matter
is assumed to be an isotropic fluid.

\begin{figure}
\centering\includegraphics[width=2.6in]{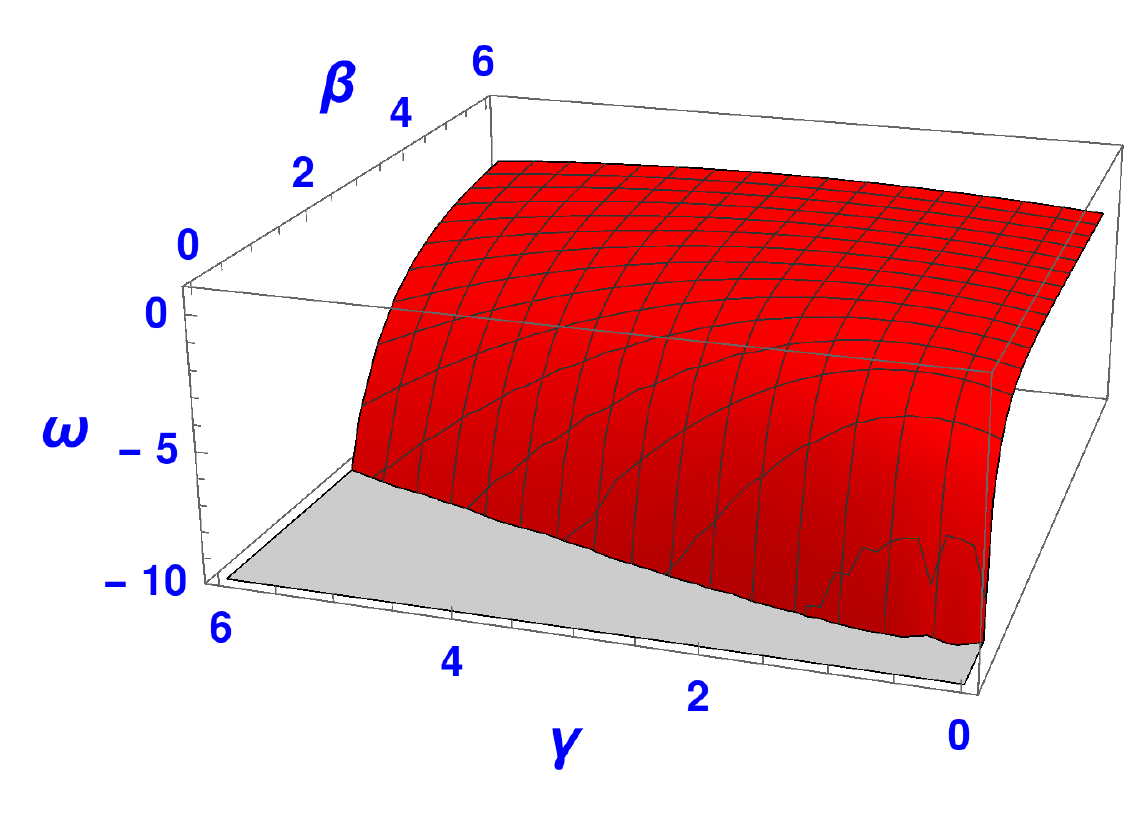}
\centering\includegraphics[width=2.6in]{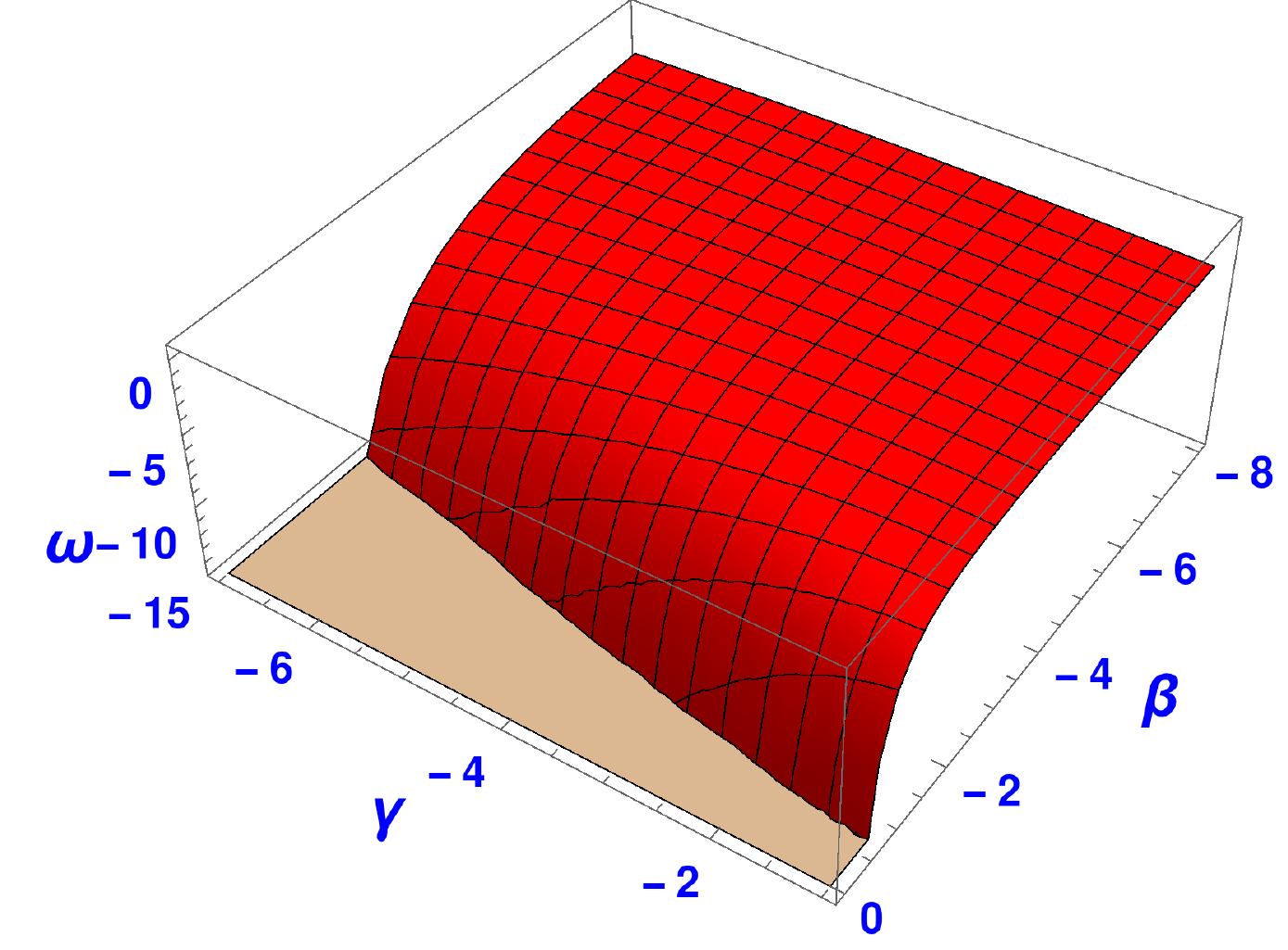}
\caption{ The left and right panels demonstrate the
allowed ranges of $\omega$, associated with the exponential-law
solution, in terms of $\gamma$ and $\beta$ for the cases (i) and (ii), respectively.
These figures show that for the allowed ranges of $\gamma$ and $\beta$,
the BD coupling parameter always takes negative values.}
\label{omega-BI-exp-law}
\end{figure}

In order to get an expanding universe, from the relation
associated with $V_s$ in (\ref{phys.quant-BT1-dep-exp}), by assuming that $a_0,\phi_0>0$, we get two
classes of solutions with (i) $b_0<0,\gamma>-1$ and (ii) $b_0>0$, $\gamma<-1$.
Let us restrict ourselves to the physical solutions.
More concretely, we would consider the following assumptions:
as the weak energy condition must be satisfied, the induced
energy density should decrease with cosmic time; the fifth dimension should be contracted when cosmic time grows~\cite{OW97}.
Consequently, for the cases (i) and (ii), we get $0<\beta<\gamma+1$
and $\gamma+1<\beta<0$, respectively. For both the cases (i) and (ii),
 the BD coupling parameter is restricted as $\omega<(\gamma+1-2\beta)/\beta$.
 According to relation (\ref{omega-exp}) and assuming the
obtained constraints on $\beta$ and $\gamma$, our numerical analysis show
that the BD coupling parameter always takes negative values, see figure \ref{omega-BI-exp-law}.
 We should note that if we do not restrict ourselves to get a contracting fifth
 dimension, then there is no upper (lower) bound for $\beta$ as $\gamma+1$.
 Moreover, with this assumption, the constraint on the BD coupling
 constant is replaced by a more generalized one. Therefore, we can obtain a much wider set of (extended) solutions.

\begin{figure}
\centering\includegraphics[width=2.8in]{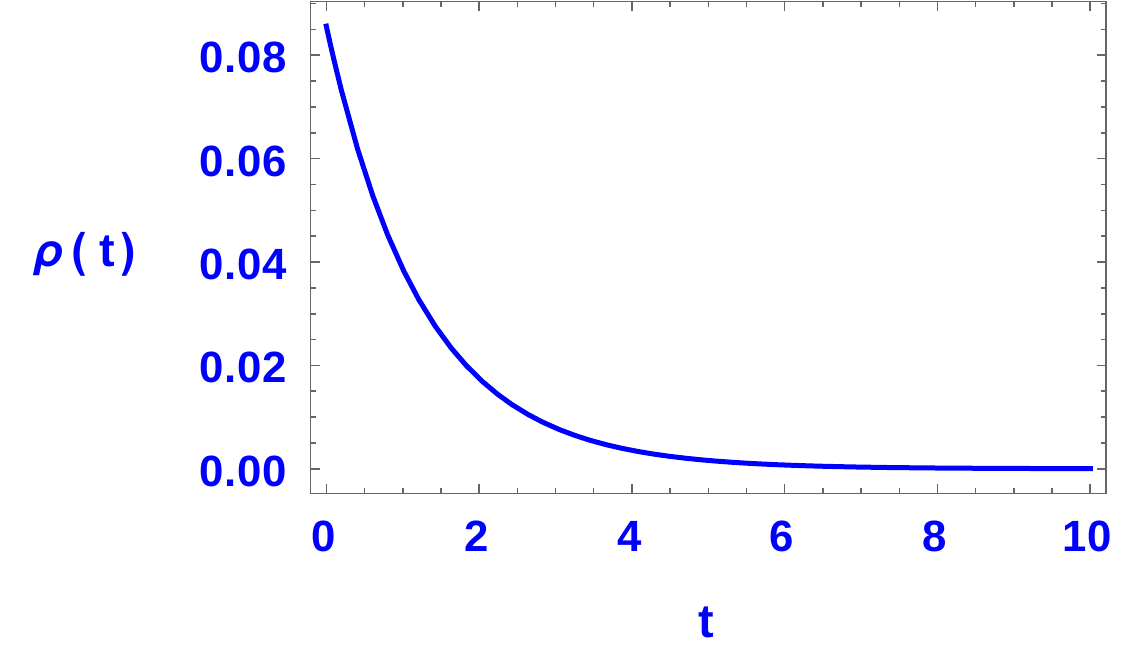}\hspace{4mm}
\centering\includegraphics[width=2.8in]{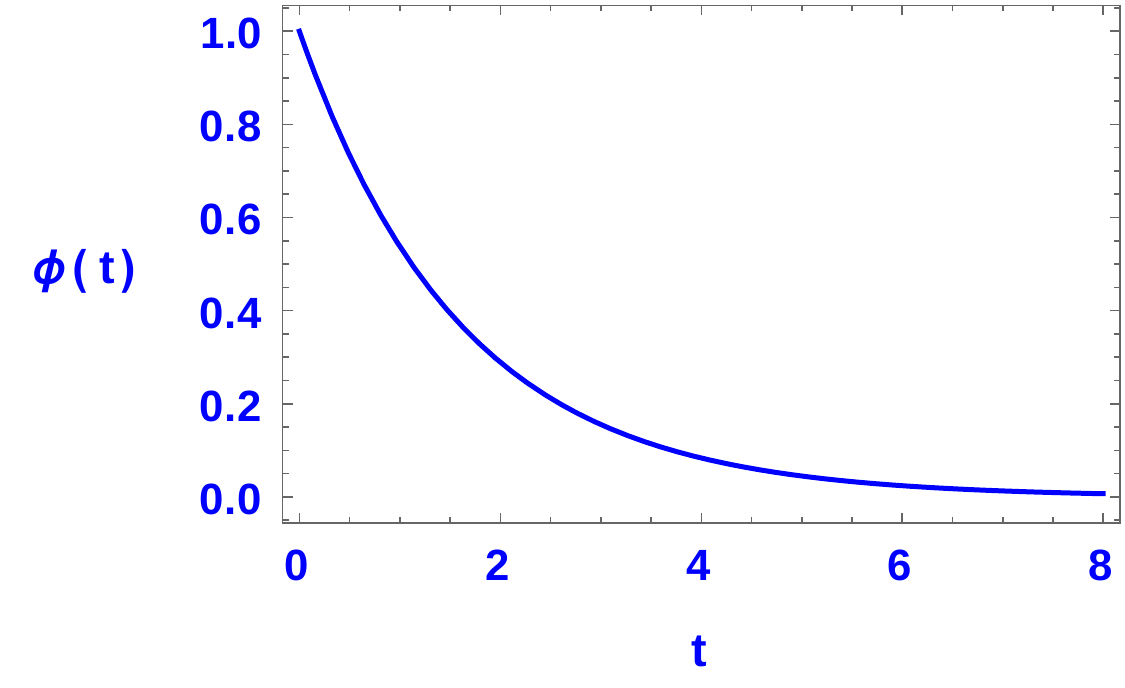}\hspace{4mm}
\centering\includegraphics[width=2.8in]{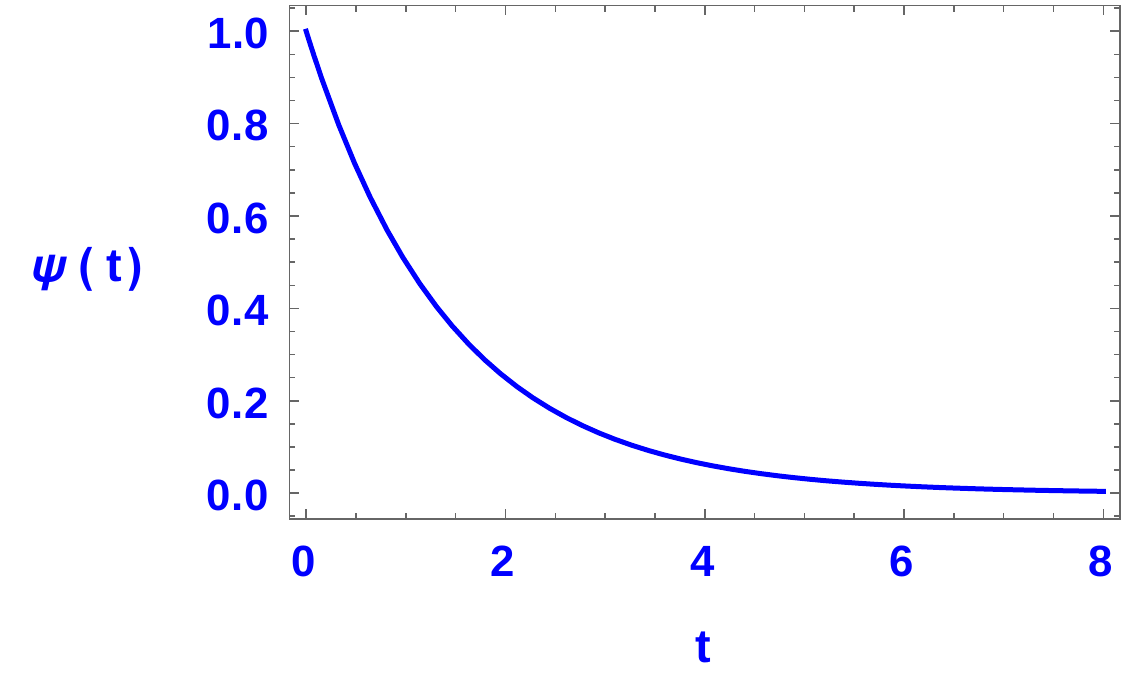}\hspace{4mm}
\centering\includegraphics[width=2.8in]{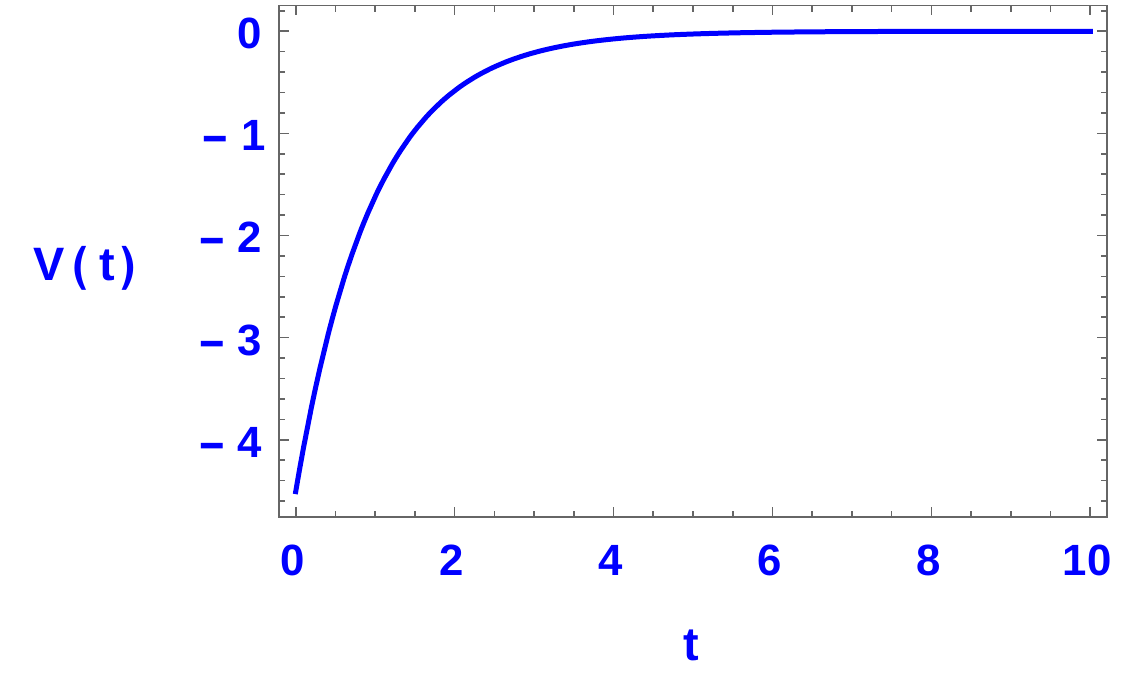}\hspace{4mm}
\caption{ The time behaviors of the induced energy
density, BD scalar field, the scalar field associated with the fifth dimension and induced scalar potential
associated with whether the case (i) or case (ii).
Upper left panel: $b_0=-1.43$, $\beta=1.1362$ and $\gamma=1.24$ are associated with case (i).
Upper right panel: $b_0=3.78$, $\beta=-2.33$ are associated with case (ii).
Lower left panel: $b_0=0.17$, $\beta=-1.15624$ and $\gamma=-2.4$ are associated with case (ii).
Lower right panel: $b_0=-1.8$, $\beta=1.84204$ and $\gamma=2.6$ are associated with case (i).
We have set other constants, such as $\phi_0$,$\eta_0$ and $\psi_0$, equal to one.}
\label{rho-phi-psi-v}
\end{figure}

Let us summarize the properties of the solutions. The above constraints on the
parameters as well as integration constants, for both of the cases (i) and (ii), provide
an exponentially expanding universe with the following properties. (i) From relations (\ref{phys.quant-BT1-dep-exp}), we get
$\frac{\sigma^2}{\theta^2}=\frac{1}{3}\left(\frac{2-\gamma}{\gamma+1}\right)^2={\rm constant}$, which
indicates that the model, in general, does not near isotropy when the cosmic time takes large values.
(ii) The average Hubble, mean anisotropy, deceleration, scalar and shear
scalar parameters always take constant values.
(iii)  The volume $V_s$ starts its exponential expansion from a nonzero constant.
(iv) The induced energy density, the BD scalar field and
and the fifth dimension decrease while the cosmic time increases.
Moreover, they tend to zero when the cosmic time takes very large values.
As a few examples, we have plotted their behaviors in terms of cosmic time in figure \ref{rho-phi-psi-v}.
(v) The induced scalar potential always increases with cosmic
time, and it tends to zero when the cosmic time takes large values.


Regarding the isotropic fluid, by setting $\delta_e=0$ in (\ref{exp-t-11-BT1-delta}) and using (\ref{omega-exp}), we
get either $\gamma=2$ or $\beta=0$. In what follows, let us discuss their
 corresponding solutions.

\subsubsection{$\beta=0$}
In the limit $\beta\rightarrow0$, we find that $|\omega|\rightarrow\infty$, and
consequently, the BD scalar field takes a constant value.
In this case, the field equations (\ref{dot-1})-(\ref{dot-5}) are satisfied
just for $\gamma=\pm i\sqrt{2}$ (where $i^2=-1$), which is not of interest.
This consequence may suggest that
 the general relativistic solutions are not always recovered from the
 corresponding BD solutions in the particular case where $\omega$ goes to infinity~\cite{BR93,BS97,Faraoni99}.

\subsubsection{$\gamma=2$}
By substituting this value for $\gamma$ into the solutions associated with the exponential-law, we obtain
\begin{eqnarray}\label{iso-metric-exp}
ds^{2}\!&=&\!-dt^{2}+\left(\frac{b_0}{\eta_0}\right)^2{\rm exp}\left(2H_0t\right)
\left[dr^2+d\Omega^2_{\zeta=0}\right],\\\nonumber
\\\nonumber
\\
\label{iso-V-exp}
V(t)\!&=&\!-\frac{2}{b_0^2}[\beta(\beta-3)(1+\omega)]\phi(t),
\hspace{35mm} \phi(t)=\phi_0\eta_0^\beta\,{\rm exp}\left(-\beta H_0 t\right),
\\\nonumber
\\
\label{iso-rho-exp}
\psi(t)\!&=&\!\psi_0\eta_0^{3-\beta}{\rm exp}\left[(\beta-3)H_0t\right], \hspace{41mm}\omega=-\frac{2 \left(\beta ^2-3 \beta+6\right)}{\beta^2},
\\\nonumber
\\
\label{iso-w-exp}
\rho(t)\!&=&\!\frac{\phi_0\eta_0^\beta}{8\pi b_0^2}(\beta-3) \left[\beta(\omega+2)-3\right]{\rm exp}\left(-\beta H_0t\right), \hspace{12mm} w=-\frac{\beta(\omega+1)+1}{\beta(\omega+2)-3},
\hspace{30mm},
\end{eqnarray}
where $H_0\equiv-1/b_0$, $A_h=0=\sigma$ and we have assumed $a_0=b_0$. These results show
a homogenous and isotropic spatially flat
 FRW universe in four dimensions. We should note that, in the context
 of the BD theory (with or without an ad hoc scalar potential), to the best
 of our knowledge, this set of solutions seems entirely novel and nobody has obtained them yet.
 It is seen that an exponentially accelerating universe can be obtained by assuming $\eta_0,b_0<0$.
 Concerning this case, let us focus on a particular case
 where $\omega=-4/3$ (for other interesting cases as $\omega=-1,0$, our model
 does not yield appropriate physical solutions).
 From (\ref{iso-w-exp}), we get $\beta=3,6$. Let us investigate the solutions associated with these particular cases.

\begin{itemize}
  \item

 For $\beta=3$, the fifth dimension
  takes constant value, the components of the effective matter as
  well as the induced scalar potential vanish.
  Therefore, we get a vacuum spatially flat FRW-BD universe
  \begin{eqnarray}\label{iso-metric-exp-vac}
ds^{2}=-dt^{2}+\left(\frac{b_0}{\eta_0}\right)^2{\rm exp}\left(2H_0t\right)
\left[dr^2+d\Omega^2_{\zeta=0}\right],\hspace{6mm}\phi(t)=\phi_0\eta_0^3\,{\rm exp}\left(-3H_0t\right),
\end{eqnarray}
  which is exactly the Ohanlon-Tupper
  solution~\cite{Faraoni.book,o'hanlon-tupper-72,KE95,MW95} for $\omega=-4/3$ in the context of the standard BD theory.
 As claimed in \cite{Faraoni.book}, this is the
  only de Sitter solution associated with the vacuum spatially flat
  universe in the standard BD theory with vanishing scalar potential.

\item
  For $\beta=6$, we get $w=1$ which corresponds to the stiff fluid.
  Moreover, from relations (\ref{iso-metric-exp})-(\ref{iso-w-exp}), we write
  \begin{eqnarray}\label{iso-metric-exp-stiff}
ds^{2}\!&=&\!-dt^{2}+\left(\frac{b_0}{\eta_0}\right)^2{\rm exp}\left(2H_0t\right)
\left[dr^2+d\Omega^2_{\zeta=0}\right],\\\nonumber
\\\nonumber
\\
\label{iso-V-exp-stiff}
V(t)\!&=&\!\frac{12\phi(t)}{b_0^2}, \hspace{37mm} \phi(t)=\phi_0\eta_0^6\,{\rm exp}\left(-6 H_0 t\right),
\\\nonumber
\\
\label{iso-rho-exp-stiff}
\psi(t)\!&=&\!\frac{\psi_0}{\eta_0^3}{\rm exp}\left(3H_0t\right), \hspace{25mm}
\rho(t)=\frac{3\phi_0\eta_0^6}{8\pi b_0^2}\,\,{\rm exp}\left(-6 H_0t\right).\hspace{15mm}
\end{eqnarray}
For $b_0<0$, we get an exponentially accelerating unverse.
Moreover, by assuming $\phi_0>0$, the BD scalar field, the induced
energy density and scalar potential decrease exponentially with cosmic time.
However, by assuming $\frac{\psi_0}{\eta_0^3}>0$, we see that the fifth dimension increases with cosmic time.
\end{itemize}

\section{Conclusions}


\label{conclusion}
In this paper, by considering an extended version of Kantowski-Sachs,
LRS Bianchi type I and Bianchi type III models as a background
space-time, we have investigated the $4D$ cosmologies that can be extracted within the MBDT \cite{RFM14}.
In this respect, we first solved the extended equations of motion in a $5D$ bulk in vacuum.
In order to solve these non-linear coupled differential equations,
 we defined a new time
coordinate, which appropriately transforms the system of field
equations into more easier counterparts.
Consequently, we have obtained new exact solutions associated with each spatial curvature in the bulk.
Moreover, for all solutions, a set of parameters and integration constants was extracted.
We have shown that they are not independent, and found
constraints which relate them to each other and to the BD coupling parameter.
Subsequently, we analyzed the solutions by discussing the allowed values
of the corresponding parameters. Moreover, by considering a few particular cases, we
have compared our results with corresponding solutions
obtained in the conventional BD theory and in general relativity.

The main objective of this paper was to discuss the reduced cosmologies
on a $4D$ hypersurface, produced from applying the methodology of the MBDT.
Therefore, we obtained expressions for induced
 physical quantities such as spatial volume, average Hubble
parameter, mean anisotropy parameter, the deceleration parameter
and the expansions for scalar expansion and the shear scalar.
Moreover, we presented the properties and behaviors of
these quantities, and discussing them.
Concerning the solutions associated with LRS Bianchi type I model, we have also obtained
all the induced physical quantities in terms of the cosmic time.

Concerning the Bianchi type I model, the scope of our solutions
are more extended than those obtained in previous investigations, see for instance \cite{RFS11}.
We have shown that for $\zeta=0$, there are two
general classes of anisotropic solutions.
The first class is power-law in terms of the cosmic time, which have
been also widely investigated in the context of general relativity, standard BD
 theory, MBDT as well as the generalized scalar tensor theories \cite{RFS11,PW08,LP-rev}.
 However, for the sake of
 completeness and comparison,
 we have analysed them
 briefly in this paper. Furthermore, we obtained new exact solutions for particular
 values of the BD coupling parameter as well as for the equation of state parameter.
 We have addressed the physical quantities which are important for
 both anisotropic and isotropic fluids.
 It should be noted that among the isotropic reduced cosmologies presented
 in this paper, the consequences associated with the stiff fluid are completely
 new and have not been investigated in the previous publications
  associated with the MBDT, see for instance, \cite{Ponce1,RFM14,RM16}, and references therein.
Moreover, the second class (which is exponential-law in terms of the cosmic time) to
 the best of our knowledge has not been obtained in the corresponding standard models as well
 as in the context of the MBDT, see, for instance, \cite{RFS11}, and references therein.
 We investigated this class comprehensively in this work regarding the anisotropic and isotropic cosmologies.
 We have also presented solutions for particular well known
 values of the BD coupling parameter as well as equation of state
 parameter including stiff, radiative and false vacuum fluids.
 For each solution of the Bianchi type I model, we have also
described the evolution of the extra
dimension (in terms of the cosmic time). Let us also mention those pointed
in \cite{BG18-2}, albeit obtained within another physical context.

\section*{Acknowledgments}
We thank the anonymous referees for their valuable comments.
S. M. M. Rasouli is grateful for the support of
Grant No. SFRH/BPD/82479/2011 from the Portuguese
Agency Funda\c c\~{a}o para a Ci\^encia e Tecnologia.
PVM is grateful to DAMTP for hospitality during his
sabbatical. This research work was supported by
Grant No. UID/MAT/00212/2019 and COST Action CA15117 (CANTATA).\\


\end{document}